\documentclass[lineno]{jfm}

\usepackage{graphicx}
\usepackage{newtxtext}
\usepackage{newtxmath}
\usepackage{natbib}
\usepackage{hyperref}
\hypersetup{
    colorlinks = true,
    urlcolor   = blue,
    citecolor  = black,
}

\newcommand{\RomanNumeralCaps}[1]
\linenumbers

\usepackage{subcaption}
\usepackage{float}

\title{Nonlocality of Mean Scalar Transport in Two-Dimensional Rayleigh-Taylor Instability Using the Macroscopic Forcing Method}

\author{D. L. O.-L. Lavacot\aff{1, 2}
  \corresp{\email{dlol@stanford.edu}},
  J. Liu\aff{1},
  H. Williams\aff{1, 2},
  B. E. Morgan\aff{2},
  \and A. Mani\aff{1}
  }

\affiliation{\aff{1}Department of Mechanical Engineering, Stanford University, Stanford, CA
\aff{2}Lawrence Livermore National Laboratory, Livermore, CA}

\begin{document}

\maketitle

\begin{abstract}
The importance of nonlocality of mean scalar transport in 2D Rayleigh-Taylor Instability (RTI) is investigated.
The Macroscopic Forcing Method (MFM) is utilized to measure spatio-temporal moments of the eddy diffusivity kernel representing passive scalar transport in the ensemble averaged fields.
Presented in this work are several studies assessing the importance of the higher-order moments of the eddy diffusivity, which contain information about nonlocality, in models for RTI.
First, it is demonstrated through a comparison of leading-order models that a purely local eddy diffusivity is insufficient in capturing the mean field evolution of the mass fraction in RTI.
Therefore, higher-order moments of the eddy diffusivity operator are not negligible.
Models are then constructed by utilizing the measured higher-order moments. 
It is demonstrated that an explicit {operator} based on the Kramers-Moyal expansion of the eddy diffusivity kernel is insufficient. 
An implicit {operator} construction that matches the measured moments is shown to offer improvements relative to the local model in a converging fashion.
\end{abstract}

\begin{keywords}
Authors should not enter keywords on the manuscript, as these must be chosen by the author during the online submission process and will then be added during the typesetting process (see http://journals.cambridge.org/data/\linebreak[3]relatedlink/jfm-\linebreak[3]keywords.pdf for the full list)
\end{keywords}

\section{Introduction}
Rayleigh-Taylor Instability (RTI) is a phenomenon that occurs when a heavy fluid is accelerated into a light fluid.
Specifically, RTI occurs when the following are present: 1) a density gradient, 2) an acceleration (associated with the body force) in the direction opposite that of the density gradient, and 3) a perturbation at the interface of the two fluids.
RTI is present in many scientific and engineering applications such as supernovae \citep{gull1975} and inertial confinement fusion (ICF) \citep{zhou2017,lindl1995}.
In the case of ICF, RTI occurs when a perturbation forms between the outer heavy ablator and the inner light deuterium gas, which causes premature mixing in the target, thereby greatly reducing the efficiency of the process.
Thus, RTI is of great interest to scientists and engineers, especially in the context of ICF.

During a typical ICF experiment design process, a Reynolds-Averaged Navier-Stokes (RANS) approach is often utilized to model the role of hydrodynamic instabilities such as RTI.
This is despite the fact that RTI can be more accurately predicted using high-fidelity methods like direct numerical simulations (DNS) \citep{youngs1994, cookdimotakis2001, cookzhou2002, cabotcook2006, mueschkeschilling2009} and large eddy simulations (LES) \citep{darlington2002, cookcabotmiller2004, cabot2006}.
Motivation for development of RANS models for various engineering applications like ICF can be understood by considering the computational cost of each method.
DNS requires resolution of the smallest turbulent scales, and LES the energy-containing scales, which are still much smaller than the macroscopic physics (i.e., averaged fields) of engineering interest.
On the other hand, by design, RANS must only resolve macroscopic scales, thereby requiring much lower computational cost.
Thus, RANS models are commonly used in engineering practice, especially in design optimization, where hundreds of thousands of simulations are often performed.
Such is especially the case in designing targets for ICF experiments \citep{casey2014, khan2016}.
Due to the utility of RANS in such applications, the need for predictive RANS models remains salient.

Models of varying complexities have been applied to the RTI problem.
Among the most commonly-used types are two-equation models.
One such model is the ubiquitously-used $k$-$\varepsilon$ model \citep{launder1974}.
Particularly, \citet{gauthierbonnet1990} introduced algebraic relations for some closures to satisfy realizability constraints for the model to be valid under the strong gradients of RTI.
Another popular two-equation model is the $k$-$L$ model; a version was introduced by \citet{dimontetipton2006} for RTI.
One appeal of the $k$-$L$ model is its inclusion of a transport equation for turbulence lengthscale $L$ (in place of the transport equation for $\varepsilon$ in $k$-$\varepsilon$) that can be related to the initial interface perturbation.
The self-similarity of turbulent RTI is leveraged to set the model coefficients.

These two-equation models rely on the gradient diffusion approximation for the turbulent mass flux closure.
The gradient diffusion approximation rests on the assumption that turbulence transports quantities in a manner similar to Fickian diffusion.
Importantly, this approximation implies purely local dependence of the mean turbulent flux on the mean gradient, ignoring history effects and gradients at nearby points in space.
However, this approximation may not be valid for mean scalar transport.
Specifically, the turbulent mass flux contains features that the gradient diffusion approximation cannot capture \citep{morgangreenough2015, denissen2014}, so a local coefficient may not be enough to scale the mean gradient to model turbulent mass flux.

Nonlocality in RTI has been studied in experiments and simulations.
\citet{clark1997} analyzed data from turbulent RTI experiments and compared the pressure-strain correlation and pressure production due to turbulent mass flux, suggesting spatial nonlocality of pressure effects.
DNS studies by \citet{ristorcelliclark2004} and experiments by \citet{mueschke2006} have also examined nonlocality of RTI in the context of two-point correlations.
Thus, the nonlocal nature of RTI is well-known, and work has been done to capture this nonlocality in models.
{
For example, two-point closures to account for nonlocality in RTI have been developed by several authors for RANS  \citep{clark1995two, steinkamp_i_1999, steinkamp_ii_1999, 
pal2018, kurien2022local}  and LES \citep{parishduraisamy2017}.
}
While these works attempt to address the effects of nonlocality in RTI, they do so without directly studying the form of the nonlocal operator.

Several authors have studied ways to directly measure the nonlocal eddy diffusivity in other canonical flows.
One such approach involves application of the Green's function.
The Green's function approach starts from analytical derivations of relations between turbulent fluxes and mean gradients, which was done by \citet{kraichnan1987}.
\cite{hamba1995} then introduced a reformulation of these relations appropriate for numerical computation of nonlocal eddy diffusivities, which has been applied to study channel flow \citep{hamba_2004} and, most recently, homogeneous isotropic turbulence (HIT) \citep{hamba2022}.

A different approach to determining nonlocal eddy diffusivities is the Macroscopic Forcing Method (MFM) by \citet{manipark2021}.
In contrast to the Green's function approach, MFM is derived by considering arbitrary forcing added directly to the transport equations with its formulation rooted in linear algebra.
Additionally, MFM offers extensions to the Green's function approach by utilization of forcing functions that are not of the form of a Dirac delta.
Harmonic forcing has been utilized to derive analytical fits to nonlocal operators in Fourier space \citep{shirian2022}.
Additionally, forcing polynomial mean fields using inverse MFM offers a computationally economical path for determination of spatio-temporal moments of the eddy diffusivity operator in conjunction with the Kramers-Moyal expansion as opposed to computation of the moments from a full MFM analysis through post-processing \citep{manipark2021}.
{
Previous works using MFM have revealed turbulence operators for a variety of flows.
\citet{shirian2022} and \citet{shirian2022b} measured nonlocal operators in space and time in HIT.
Though the spatial nonlocal operator was measured in HIT, it was applied to a turbulent round jet and was shown to match experiments more closely than the purely local Prandtl mixing-length model.
MFM has also been applied to turbulent wall-bounded flows, including channel flow \citep{park2023a} and separated boundary layers \citep{parkliumani2022, park2023b}, to measure the anisotropic but local eddy diffusivity.
In those flows, incorporation of the MFM-measured anisotropic eddy diffusivity improved RANS model predictions significantly, and remaining model errors were attributed to missing nonlocal effects. 
}

It is with a motivation towards RANS model improvement that the present work seeks to understand nonlocality of closure operators governing turbulent scalar flux transport in RTI using MFM.
{Note that it is not intended for MFM to supplant current RANS models.
Instead, MFM is an analysis tool that can be used to assess models and discover the necessary characteristics for accurate models.
Here, MFM allows for direct measurement of nonlocal closure operators, which has not yet been done in RTI.}
This new knowledge of nonlocality of the mean scalar transport closure operator in RTI will aid in the development of improved RANS models used for studying ICF.

{It is important to note that this work presents MFM measurements for a simplified RTI problem: the flow is two-dimensional, incompressible, and low-Atwood number, and only passive scalar mixing is considered.
Since the eddy diffusivity is not universal, the MFM measurements of its moments presented here cannot be directly extended to more complex RTI.
However, valuable insight into trends in the eddy diffusivity for mean scalar transport in RTI can be gained in this work.
This follows the common process for developing turbulence models, where models are first designed for simpler flows then tested on and adjusted for more complex flows.
In this work, MFM is performed on a simplified RTI problem to give a preliminary look into the eddy diffusivity of Rayleigh-Taylor-type flows, but future work will involve extensions to more complex flow characteristics that are closer to the practical flow observed in ICF capsules.
The intent of this work is to present MFM as a tool for determining characteristics of the eddy diffusivity of a flow (i.e., its nonlocality and the importance of its higher order moments) that a model should satisfy in order to accurately predict mean scalar transport.
The current work will inform future studies with additional complexities, including three-dimensionality, finite Atwood number, compressibility, and coupling with momentum.}

This work is organized as follows.
First, an overview of RTI is covered briefly in \S \ref{sec:rti_physics}.
Next, \S \ref{sec:math_methods} gives an overview of the mathematical methods used in this work, including: 1) the generalized eddy diffusivity and its approximation via a Karmers-Moyal expansion; 2) MFM and its application for finding the eddy diffusivity moments; 3) self-similarity analysis.
Simulation details, including the governing equations and the computational approach, are given in \S \ref{sec:sim_details}.
Finally, results of several studies on the importance of higher-order eddy diffusivity moments as well as assessments of suggested {operator} forms incorporating nonlocality of the eddy diffusivity for mean scalar transport in RTI are presented in \S \ref{sec:results}.
{The results show that nonlocality of the eddy diffusivity is important in mean scalar transport of the RTI problem studied here, and RANS models incorporating this nonlocality result in more accurate predictions than leading-order models.}

\section{Brief overview of RTI}
\label{sec:rti_physics}
RTI is characterized by spikes (heavy fluid moving into light fluid) and bubbles (light fluid into heavy fluid).
The mixing widths of these spikes and bubbles are denoted as $h_s$ and $h_b$, respectively, and the mixing half-width is defined as $h=\frac{1}{2}(h_s+h_b)$.
The behaviors of these quantities in RTI are dependent on 
the Atwood number, defined as
\begin{align}
    A=\frac{\rho_H-\rho_L}{\rho_H+\rho_L}.
\end{align}
Here, $\rho_H$ and $\rho_L$ are the densities of the heavy and light fluids, respectively.
In the limit of low-Atwood number and late time, the mixing layer width is expected to reach a self-similar state of growth that scales quadratically with time:
\begin{align}
    h\approx\alpha Agt^2,
\end{align}
where $\alpha$ is the mixing width growth rate.
The mixing width growth rate can also be viewed as the net mass flux through the midplane \citep{cookcabotmiller2004}.
In this case, $\alpha$ can also be written as
{
\begin{align}
    \alpha = \frac{\dot{h}^2}{4Agh},
\end{align}
}
where $\dot{h}$ is the time derivative of $h$.
In the limit of self-similarity, these two definitions of $\alpha$ are expected to converge to the same value.

In a simulation, $h$ can be measured as
\begin{align}
    h \equiv 4\int\langle Y_H\left(1-Y_H\right) \rangle dy,
\label{eq:h_meas}
\end{align}
where $Y_H$ is the mass fraction of the heavy fluid (therefore, $Y_L=1-Y_H$ is the mass fraction of the light fluid){, and $\langle * \rangle$ denotes averaging over realizations and homogeneous direction $x$.}
An alternative definition used in works such as \citet{cabotcook2006} and \citet{morgan2017} is
\begin{align}
    h_\text{hom} \equiv 4\int\langle Y_H \rangle \left(1-\langle Y_H \rangle \right)dy.
    \label{eq:h_hom}
\end{align}
This definition is particularly useful, since it allows $h$ to be determined solely based on the RANS field.
That is, there is no closure problem in determining $h$ with this definition.
Thus, this is the $h$ reported in this work.

From these two definitions, a mixedness parameter $\phi$ can be defined, which can be interpreted as the ratio of mixed to entrained fluid \citep{youngs1994, morgan2017}:
\begin{align}
    \phi \equiv \frac{h}{h_\text{hom}}=1-4\frac{\int\langle Y_H'Y_H' \rangle dy}{h_\text{hom}}.
\end{align}
In the limit of self-similarity, $\phi$ is expected to approach a steady-state value.

A metric for turbulent transition is the Taylor Reynolds number:
\begin{align}
    Re_T = \frac{k^{1/2}\lambda}{\nu},
\end{align}
where $k=\frac{1}{2}\langle u_i'u_i' \rangle $ is the turbulence kinetic energy, and $\lambda$ is the effective Taylor microscale, approximated by
\begin{align}
    \lambda = \sqrt{\frac{10\nu L}{k^{1/2}}}.
    \label{eq:ReT}
\end{align}
Here, the turbulent lengthscale $L$ can be approximated as $\frac{1}{5}$ the mixing layer width \citep{morgan2017}.
The large-scale Reynolds number can also be examined \citep{cabotcook2006}:
\begin{align}
    Re_L = \frac{h_{99}\dot{h}_{99}}{\nu},
\end{align}
where $h_{99}$ is the mixing width based on $1\%$-$99\%$ mass fraction.
\citet{dimotakis2000} determined that the criterion for turbulent transition is when $Re_T>100$ or $Re_L>10,000$.

\section{Mathematical methods}
\label{sec:math_methods}

\subsection{Model problem}

In this work, a two-dimensional (2D), nonreacting flow with two species---a heavy fluid over a light fluid---is considered, with gravity pointing in the negative $y$-direction. 
It must be noted that the behavior of 2D RTI is significantly different from three-dimensional (3D) RTI, the latter of which is more relevant to problems of engineering interest.
It is well known that while 2D RTI is unsteady and chaotic, it is not strictly turbulent, since turbulence is a characteristic of 3D flows.
In addition, 2D RTI has a faster late-time growth rate, develops larger structures, and is ultimately less well-mixed.
These differences have been studied in RTI by \citet{cabot2006} and \citet{young_tufo_dubey_rosner_2001} and in Richtmeyer-Meshkov instability by \citet{olson2014}.

For this study, 2D RTI is chosen as the model problem instead of 3D RTI, since it is a good simplified setting for understanding nonlocality in RTI through the lens of MFM.
Specifically, 2D RTI simulations are much less computationally expensive than those of 3D RTI, and MFM requires many simulations to attain statistical convergence.
Thus, 2D RTI remains the focus of this work, with the hope that the understanding of nonlocality in this flow could be extended to nonlocality in 3D RTI.

In this 2D problem, $x$ is the homogenous direction.
In addition, there is no surface tension, the Atwood and Mach ($Ma$) numbers are finite but small, and the Peclet ($Pe$) number is finite but large.

\subsection{Generalized eddy diffusivity and higher-order moments}
In this work, the effect of nonlocality on mean scalar transport is of interest, so analysis begins with the scalar transport equation under the assumption of incompressibility:
\begin{align}
    \frac{\partial Y_H}{\partial t} + \nabla\cdot(\mathbf{u}Y_H) = D_H\nabla^2Y_H,
    \label{eq:ste}
\end{align}
where $\mathbf{u}$ is the velocity vector and $D_H$ is the molecular diffusivity of the heavy fluid.

{After Reynolds decomposition and averaging, this becomes
\begin{align}
    \frac{\partial \langle Y_H \rangle }{\partial t} + \nabla\cdot(\langle{\mathbf{u}}\rangle\langle Y_H \rangle ) = -\frac{\partial\langle v'Y_H' \rangle }{\partial y}
    + D_H\nabla^2\langle Y_H \rangle .
    \label{eq:ste_avg_noassumptions}
\end{align}}

{In this work, large $Pe$ (the ratio of advective transport rate to diffusive transport rate) and small $A$ are assumed.
The former assumption means molecular diffusion is negligible, and the latter yields $\langle u_i \rangle =0$, allowing the advective term to drop.
Equation \ref{eq:ste_avg_noassumptions} becomes}
\begin{align}
    \frac{\partial \langle Y_H \rangle }{\partial t} = -\frac{\partial\langle v'Y_H' \rangle }{\partial y}
    \label{eq:ste_avg}.
\end{align}
The term $\langle v'Y_H' \rangle $ is the turbulent scalar flux, and this is the unclosed term that needs to be modeled.

As mentioned previously, one reason the gradient diffusion approximation used to model this term is inaccurate is that it assumes locality of the eddy diffusivity.
This assumption can be removed by instead considering a generalized eddy diffusivity that is nonlocal in space and time, as demonstrated by \citet{romanof1985} and  \citet{kraichnan1987}.
For 2D RTI, such a model reduces to
\begin{align}
    -\langle v'Y_H' \rangle (y,t) = \int\int D(y,y',t,t') \left.\frac{\partial \langle Y_H \rangle }{\partial y}\right|_{y',t'} dy' dt'
    \label{eq:gen_eddy_diff_2d}.
\end{align}
Here, $y$ is the spatial coordinate in averaged space and $t$ is the time at which the turbulent scalar flux is measured,  $y'$ is all points in averaged space, and $t'$ is all points in time.
{
This definition is exact for passive scalar transport, including in the case studied in this work.
}

{
This nonlocal eddy diffusivity can also be viewed as a two-point correlation.
This was first described by \citet{taylor1922diffusion} in homogeneous turbulence. 
Through Lagrangian statistical analysis, Taylor derived the following relation between diffusivity and velocity correlations:
\begin{align}
    D_{ij} = \int_0^\infty{ \left<v_i(t)v_j(t+t')\right> dt' }.
\end{align}
Work by \citet{shende2023model} has shown that MFM recovers this Lagrangian formulation for eddy diffusivity in homogeneous flows.
It should be noted that the above definition is not valid for inhomogeneous RTI (again, the exact definition of eddy diffusivity for the studied flow is the one in equation \ref{eq:gen_eddy_diff_2d}), but the intent here is to provide another interpretation of MFM more aligned with the well-understood two-point correlations. 
}

The eddy diffusivity kernel can be approximated by Taylor-series-expanding the scalar gradient locally about $y$ and $t$, which results in the following Kramers-Moyal-like expansion for the turbulent scalar flux as done by \citet{kraichnan1987}, \citet{hamba1995}, and \citet{hamba_2004}:
{\begin{align}
    -\langle v'Y_H' \rangle (y,t) = D^{00}(y,t)\frac{\partial \langle Y_H \rangle }{\partial y} +
    D^{10}(y,t)\frac{\partial^2 \langle Y_H \rangle }{\partial y^2} +
    D^{01}(y,t)\frac{\partial^2 \langle Y_H \rangle }{\partial t \partial y} +
    D^{20}(y,t)\frac{\partial^3 \langle Y_H \rangle }{\partial y^3} \dots
    \label{eq:explicit}
\end{align}
\begin{align}
    D^{00}(y,t) &= \int \int D(y,y',t,t')dy' dt', 
    \\
    D^{10}(y,t) &= \int \int (y'-y)D(y,y',t,t')dy' dt',
    \\
    D^{01}(y,t) &= \int \int (t'-t)D(y,y',t,t')dy' dt',
    \\
    D^{20}(y,t) &= \int \int \frac{(y'-y)^2}{2}D(y,y',t,t')dy' dt'.
\end{align}}
Here, $D^{mn}$ are the eddy diffusivity moments; the first index, $m$, denotes order in space, while the second, $n$, denotes order in time. 
This is the form presented in \citet{manipark2021} and \citet{liu2023}.

When the eddy diffusivity kernel is purely local, 
\begin{align}
    D(y,y',t,t')=D^{00}\delta(y-y')\delta(t-t').
\end{align}
In this case, $D^{00}$ is the only surviving moment, while all higher-order moments in space and time are zero.
Any non-zero higher-order moment therefore characterizes the nonlocality of the eddy diffusivity kernel.
Thus, this expansion implies explicitly a model form for the turbulent scalar flux that incorporates nonlocality of the eddy diffusivity.
Truncating the expansion provides an approximation of $\langle v'Y_H' \rangle $ but with the caveat that the expansion may not converge.
This will be discussed in more detail later in \S \ref{sec:exp_model_form}.

Each $D^{mn}$ provides more information about the eddy diffusivity kernel with increasing order.
For example, $D^{00}$ represents the volume of the kernel in space-time. 
The coefficient corresponding to one higher-order in space, $D^{10}$, provides information about the centroid of the kernel in space.  
$D^{20}$ contains information about the moment of inertia of the kernel in space, $D^{01}$ contains information about the centroid of the kernel in time, and so on.

\subsection{The Macroscopic Forcing Method}

\begin{figure}
    \centering
    \includegraphics[width=0.5\textwidth]{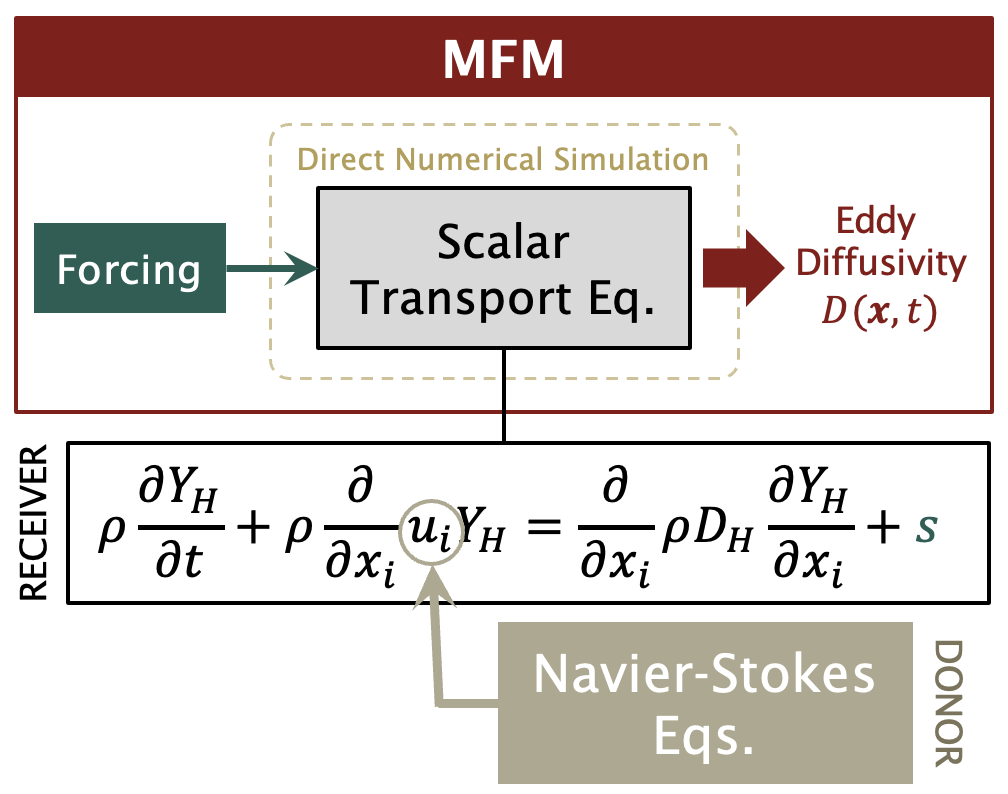}
    \caption{Diagram of MFM pipeline.}
    \label{fig:MFM_diagram}
\end{figure}

MFM is a method for numerically determining closure operators in turbulent flows \citep{manipark2021}.
Much like a rheometer measures the molecular viscosity of a fluid by imposing a shear force on the flow, MFM forces the transport equation in a turbulent flow and extracts the closure operator from its response. 
{Unlike the molecular viscosity, which is a material property, the turbulent closure operator is a property of the flow, so MFM measurements of one flow cannot be generalized for all flows; the MFM-measured closure for one flow cannot be applied exactly as it is to a different flow.
However, MFM measurements of one flow can reveal characteristics of the turbulent closure that are expected be true for a family of similar flows.

Specifically, MFM can be used to determine the RANS closure operator, as shown in the pipeline diagram in figure \ref{fig:MFM_diagram}.
In MFM, two simulations are run at once: the donor and the receiver simulation.
In this work, the donor simulation numerically solves the multicomponent Navier-Stokes equations in equations \ref{eq:gov_eq1} - \ref{eq:gov_eq2}.
The receiver simulation ``receives" $u_i$ from the donor simulation and uses it to solve the scalar transport equation with a forcing $s$:
\begin{align}
    \rho\frac{\partial Y_H}{\partial t} + \rho \frac{\partial}{\partial x_i}\left(u_iY_H\right)= \frac{\partial}{\partial x_i}\left(\rho D_H\frac{\partial}{\partial x_i}Y_H\right) + s.
    \label{eq:receiver}
\end{align}
Ultimately, forcings on the receiver simulation effect a response from the flow, and measuring this response allows for determination of the eddy diffusivity.
{
In particular, these forcings are \textit{macroscopic}.
Here, macroscopic quantities are defined as fields that are unchanged by Reynolds-averaging.
Mathematically, the macroscopic forcing is such that $s=\overline{s}$.
This macroscopic nature is crucial to the method, since it does not disturb the underlying mixing process, which allows for measurement of the closure operator without changing it.
}
For details, see \citet{manipark2021}.

In actuality, the Inverse Macroscopic Forcing Method (IMFM) is used to determine eddy diffusivity moments.
That is, instead of the forcings being chosen, certain mean mass fraction fields are chosen.
{Numerically, mean mass fractions are enforced in each realization, so the averages (denoted by $\overline{*}$) described here are in $x$, the homogeneous direction in space.}
The forcing needed to maintain the chosen $\overline{ Y_H } $ is determined implicitly along the process and is not directly used in the analysis.

As an illustration, the measurement of $D^{00}$ can be considered.
According to equation \ref{eq:explicit}, choosing $\overline{ Y_H } =y$ (for $y$ between $-1/2$ and $1/2$) results in $\frac{\partial\overline{ Y_H } }{\partial y}=1$, and all other higher-order derivatives are zero.
Thus, choosing this $\overline{ Y_H }$ {in each realization} results in {the realization- and spatially-averaged} measurement $-\langle v'Y_H' \rangle =D^{00}$.

Measurement of higher-order moments involves similar choices of $\overline{ Y_H } $ but requires information from lower order moments.
For example, measuring $D^{10}$ involves choosing $\overline{ Y_H } =y^2$, which results in $-\langle v'Y_H' \rangle =yD^{00}+D^{10}$.
Here, $D^{00}$ comes from the simulation using $\overline{ Y_H } =y$.
Thus, $D^{10}$ is computed by subtracting $yD^{00}$ from the $\langle v'Y_H' \rangle $ measurement from the simulation using $\overline{ Y_H } =y^2$.

Specifically, the following desired mean mass fractions are used for each moment for $y$ between $-1/2$ and $1/2$:
\begin{align}
    \overline{ Y_H } &=y \Rightarrow D^{00}\label{eq:des_YH_1},\\
    \overline{ Y_H } &=\frac{1}{2}y^2 \Rightarrow D^{10},\\
    \overline{ Y_H } &=yt \Rightarrow D^{01},\\
    \overline{ Y_H } &=\frac{1}{6}y^3+\frac{1}{48} \Rightarrow D^{20}\label{eq:des_YH_2}.
\end{align}
From these $\overline{ Y_H } $, the needed forcing in each timestep is numerically determined:
\begin{align}
    s^k = \frac{\overline{ Y_H } _\text{desired}^k - \overline{ Y_H } ^{k-1}}{\Delta t}
\end{align}
where the superscript $k$ denotes the timestep number, $\overline{ Y_H } _\text{desired}$ is the mean mass fraction desired as outlined in equations \ref{eq:des_YH_1} - \ref{eq:des_YH_2}, and $\Delta t$ is the timestep size.

{
This MFM forcing bears some resemblance to other forcings used in the literature, such as interaction by exchange with the mean (IEM) \citep{sawford2004conditional, pope2001turbulent}.
One main difference between forcings in such methods and MFM is that the purpose of the latter is to drive the flow to a specified mean gradient, which allows for measurement\textendash not enforcement \textendash of the eddy diffusivity moments.
In other words, in MFM for scalar transport, the input is a mean scalar gradient, and the output is the eddy diffusivity moment; in IEM and similar methods, the input is a desired moment (e.g., in IEM, the input moment is $\langle c^2\rangle$) and the output is a mixing model.
In addition, methods such as IEM use microscopic forcings, while MFM uses macroscopic forcings, which is a distinguishing characteristic of the latter method.
}

To determine $D^{00}$, $D^{10}$, $D^{01}$, and $D^{20}$, four separate simulations are needed.
For each of these simulations, the moments can be calculated using measurements of the turbulent scalar flux as follows:
\begin{align}
    D^{00}&=F^{00},\\
    D^{10}&=F^{10}-yD^{00},\\
    D^{01}&=F^{01}-tD^{00},\\
    D^{20}&=F^{20}-yD^{10}-\frac{1}{2}y^2D^{00}.
\end{align}
where $F^{mn}$ denotes the $-\langle v'Y_H' \rangle $ measured from the receiver simulation using the forcing corresponding the moment $D^{mn}$.

\subsection{Self-similarity analysis}

We perform our analysis in the self-similar regime.
First, we define a self-similar coordinate:
\begin{align}
    \eta = \frac{y}{h(t)},
\end{align}
so that $\langle Y_H \rangle $ is only a function of $\eta$.
Note that $\eta$ requires a definition of $h(t)$.
From the previous discussion on the self-similarity of RTI, an appropriate definition is $h(t)=\alpha Agt^2$.

Through self-similar analysis of equation \ref{eq:explicit}, the eddy diffusivity moments and turbulent scalar flux can be normalized.
Details of this process can be found in the Appendix.

\subsection{Algebraic fit to mixing width}
Recall that $h(t)=\alpha Agt^2$ is used in the self-similarity analysis.
This is valid only for late time, so the subsequent analyses in this work are all done in this self-similar timeframe.
Usually, $\alpha$ can be determined from $\frac{h(t)}{Agt^2}$, where $h(t)$ is computed from the simulation via equation \ref{eq:h_hom}.
However, due to the convergence and statistical errors as well as the existence of a virtual time origin, $\alpha Agt^2$ is not a good representation of $h(t)$ measured in the DNS.
Instead, a fitting coefficient $\alpha^*$ and virtual time origin $t^*$ are determined to make a shifted quadratic fit to $h(t)$ from the simulation:
\begin{align}
    h_\text{fit}(t)=\alpha^*Ag(t-t^*)^2.\label{eq:hfit}
\end{align}

With this fit, the normalizations of the turbulent scalar flux and moments become
\begin{align}
    \widehat{\langle v'Y_H' \rangle }=&\frac{\langle v'Y_H' \rangle }{\alpha^* Ag(t-t^*)}\label{eq:nondim1},\\
    \widehat{D^{00}}=&\frac{D^{00}}{{\alpha^*}^2 A^2g^2(t-t^*)^3},\label{eq:nondimD00}\\
    \widehat{D^{10}}=&\frac{D^{10}}{{\alpha^*}^3 A^3g^3(t-t^*)^5},\\
    \widehat{D^{01}}=&\frac{D^{01}}{{\alpha^*}^2 A^2g^2(t-t^*)^4},\\
    \widehat{D^{20}}=&\frac{D^{20}}{{\alpha^*}^4 A^4g^4(t-t^*)^7}\label{eq:nondim2}.
\end{align}

For exact self-similarity, plots of the measured $\widehat{D^{mn}}$ against $\eta$ must be independent of time.
This expectation sets a criterion to assess the extent to which ideal self-similarity is achieved.
Plots and assessment of the self-similar collapse of the measurements presented in this work are in the Appendix.

\section{Simulation details}
\label{sec:sim_details}

\subsection{Governing equations}
The governing equations solved in this work are the compressible multicomponent Navier-Stokes equations, which involve equations for continuity, diffusion of mass fraction $Y_\alpha$ of species $\alpha$ (characterized by its binary molecular diffusivity $D_\alpha$), momentum transport, and transport of specific internal energy $e$:
\begin{align}
    \frac{D\rho}{Dt}&=-\rho\frac{\partial u_i}{\partial x_i}\label{eq:gov_eq1},\\
    \rho\frac{DY_\alpha}{Dt}&=\frac{\partial }{\partial x_i}\left(\rho D_\alpha\frac{\partial Y_\alpha}{\partial x_i}\right),\\
    \rho\frac{Du_j}{Dt}&=-\frac{\partial }{\partial x_i}\left(p\delta_{ij}+\sigma_{ij}\right) + {\rho g_j,}\\
    \rho\frac{De}{Dt}&=-p\frac{\partial u_i}{\partial x_i}+\frac{\partial }{\partial x_i}\left(u_i\sigma_{ij}-q_j\right)\label{eq:gov_eq2}.
\end{align}
Here, $\frac{D}{Dt}$ is the material derivative $\frac{\partial}{\partial t} + u_i\frac{\partial}{\partial x_i}$, $\rho$ is density, $u$ is velocity,  $p$ is pressure, and $g$ is gravitational acceleration, active in the $-y$ direction.
The viscous stress tensor $\sigma_{ij}$ and heat flux vector $q_j$ are respectively defined as
\begin{align}
    \sigma_{ij} &= \mu\left(\frac{\partial u_i}{\partial x_j}+\frac{\partial u_j}{\partial x_i}\right)-\mu\frac{2}{3}\frac{\partial u_k}{\partial x_k}\delta_{ij},\\
    q_j &= -\kappa\frac{\partial T}{\partial x_j} - \sum^N_{\alpha=1}h_\alpha\rho D_\alpha\frac{\partial Y_\alpha}{\partial x_j}.
\end{align}
Here, $\mu$ is the dynamic viscosity, $\kappa$ is the thermal conductivity, $T$ is temperature, and $h_\alpha$ is the specific enthalpy of species $\alpha$.

Component pressures and temperatures of each species are determined using ideal gas equations of state.
Under the assumption of pressure and temperature equilibrium, an iterative process is performed to determine volume fractions $v_\alpha$ that allow for computation of partial densities and energies.
More details on the hydrodynamics equations and computation of component quantities can be found in \citet{morganolson2018}.

Finally, total pressure is determined as the weighted sum of component pressures:
\begin{align}
    p&=\sum^N_{\alpha=1}v_\alpha p_\alpha.
\end{align}

{In general, in these compressible equations, $Y_\alpha$ are not passive scalars.
However, the component equations of state are scaled so that a consistent hydrostatic pressure gradient is maintained across the mixing layer.
Thus, in this work, $Y_\alpha$ are effectively passive.}

\subsection{Computational approach}

Simulations for 2D RTI are run using the Ares code, a hydrodynamics solver developed at Lawrence Livermore National Laboratory (LLNL) \citep{morgangreenough2015, bender2021}.
Ares employs an arbitrary Lagrangian-Eulerian (ALE) method based on the one by \citet{sharp_barton_1981}, in which the governing equations (equations \ref{eq:gov_eq1} to \ref{eq:gov_eq2}) are solved in a Lagrangian frame and then remapped to an Eulerian mesh through a second-order scheme.
The spatial discretization is a second-order non-dissipative finite element method, and time advancement is a second-order explicit predictor-corrector scheme.

The Reynolds number (more specifically, the kinematic viscosity $\nu$) is set through a numerical Grashof number, such that
\begin{align}
    \nu = \sqrt{ \frac{-2gA\Delta^3}{Gr}}.
\end{align}
Here, $\Delta$ is the grid spacing; in the simulations, a uniform mesh is used, and $\Delta=\Delta x=\Delta y$.
To ensure that the unsteady structures are properly resolved and for the simulation to appropriately be considered a DNS, $Gr$ should be kept small.
A $Gr$ that is too large results in a simulation with dissipation dominated by numerics rather than the physics.
\citet{morganblack2020} found that past $Gr\approx12$ in the Ares code, numerical diffusivity dominates molecular diffusivity.
For our simulations, we use $A=0.05$ and $Gr=1$, the latter of which is in line with the DNS by \citet{cabotcook2006}.
{These choices give a $\nu$ of $10^{-9} m^2/s$.
The Schmidt number $Sc$, defined as $\nu/D_M$, is set to unity, so $D_M=10^{-9} m^2/s$.}

The Mach number, $Ma=\frac{u}{c}$, where $c$ is the speed of sound, characterizes compressibility effects of the flow.
$Ma$ is set by the specific heat ratio $\gamma$, which is $5/3$ in the simulations im this work.
The maximum $Ma$ is measured at the last timestep to be approximately $0.03$, which is ascertained to be small enough to assume incompressibility.

The Peclet number $Pe$ characterizes the advection versus diffusion rate and is defined as $ReSc$.
Here, a $Pe_L$ and a $Pe_T$ are reported, which use a large-scale $Re_L$ and the Taylor Reynolds number $Re_T$, respectively.
In the presented simulations, $Sc=1$.
The two $Pe$ are computed in post-processing: $Pe_L$ is approximately $8,000$, and $Pe_T$ is approximately $54$.
Both are below the criterion established by \citet{dimotakis2000}, suggesting that the simulated flow is transitional or pre-transitional.

The number of cells in each simulation is $2049\times2049$.
The width $L_x$ of the domain is $1$, and the height $L_y$ is $1$.
The boundary conditions are periodic in $x$ and no slip and no penetration in $y$.

Initially, the velocity field is zero, temperature is 293.15 K, and pressure is 1 atm. 
A tophat perturbation based on the ones used by \citet{morgangreenough2015} and \citet{morgan2022} is imposed on the density field at the interface of the heavy and light fluids:
\begin{align}
    \xi(x) &=
    \sum^{\kappa_\text{max}}_{k=\kappa_\text{min}}
    \frac{\Delta}{\kappa_\text{max}-\kappa_\text{min}+1}
    \left(\cos\left(2\pi k x+\phi_{1,k}\right)
    +\sin\left(2\pi k x+\phi_{2,k}\right)\right),\\
    \rho(x,y) &= \rho_L+ \frac{\rho_H-\rho_L}{2}\left(1+\tanh\left(\frac{y-L_y/2+\xi}{2\Delta}\right)\right),
\end{align}
where $\phi_{1,k}$ and $\phi_{2,k}$ are phase shift vectors randomly taken from a uniform distribution, and $L_y$ is the length in $y$ of the domain.
Here, the minimum and maximum wavenumbers are set to  $\kappa_\text{min}=8$ and $\kappa_\text{max}=256$, respectively.

The stop condition of the simulations is when $h$ is approximately half the domain size in $y$.
This corresponds to the nondimensional simulation time $\tau$ of $30.84$.
$\tau$ is defined as $\frac{t}{t_0}$, where $t_0=\sqrt{\frac{h_0}{Ag}}$ and $h_0$ is the dominant lengthscale determined by the peak of the initial perturbation spectrum.

\begin{figure}
    \centering
    \begin{subfigure}[]{0.49\textwidth}
        \includegraphics[width=\textwidth]{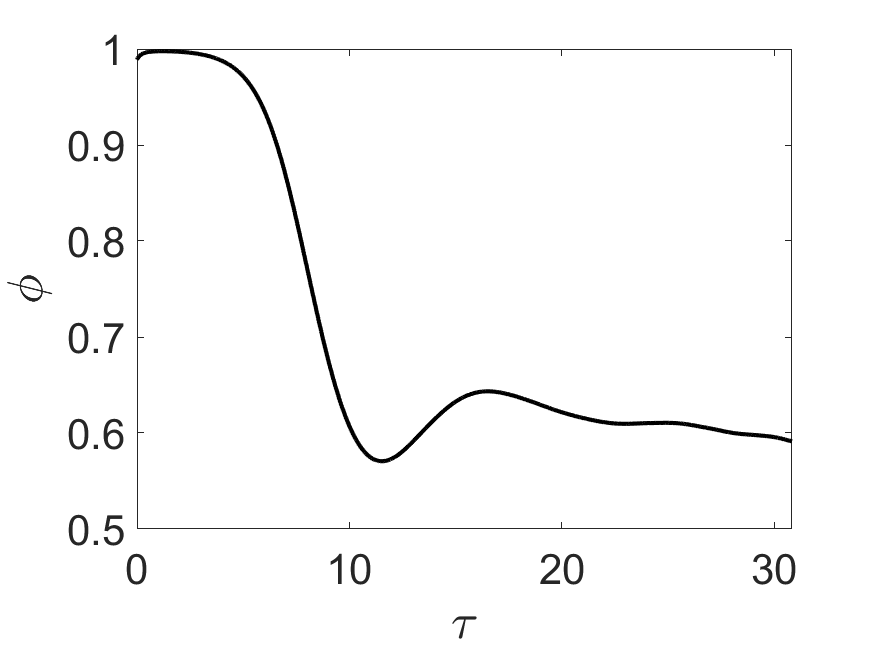}
        \subcaption[]{}
        \label{subfig:donor_phi}
    \end{subfigure}
    \begin{subfigure}[]{0.49\textwidth}
        \includegraphics[width=\textwidth]{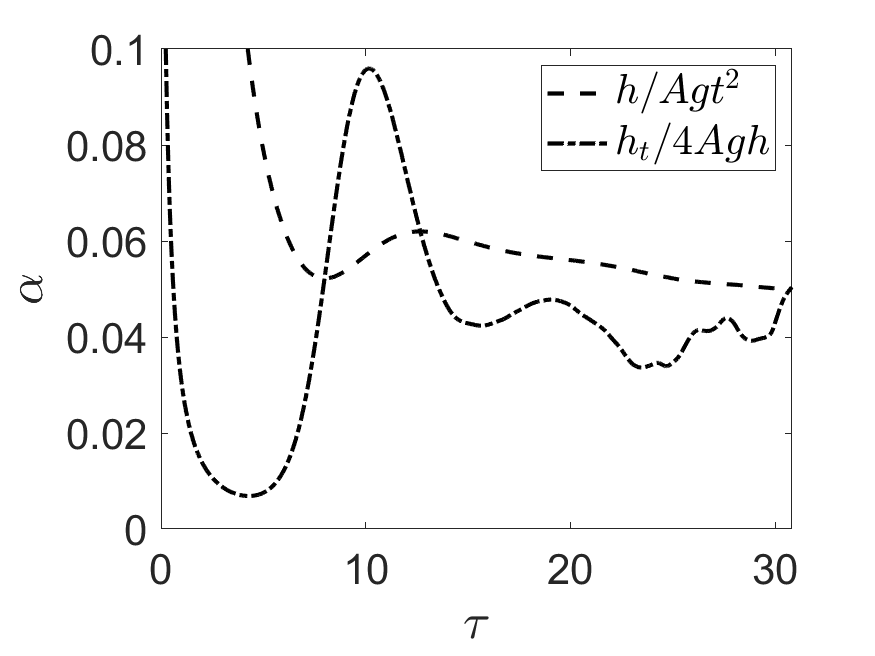}
        \subcaption[]{}
        \label{subfig:donor_alpha}
    \end{subfigure}
    \caption{Self-similarity parameters computed from a donor simulation.}
    \label{fig:donor}
\end{figure}

\begin{figure}
    \centering
    \includegraphics[width=0.49\textwidth]{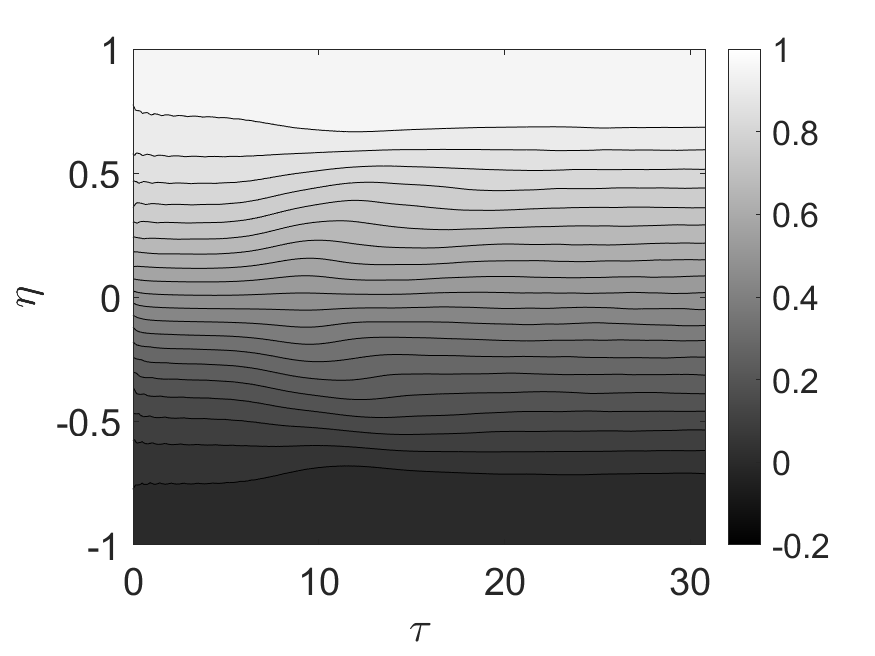}
    \caption{{Contours of $\langle Y_H \rangle $ showing self-similarity at late times.}}
    \label{fig:YH_contour}
\end{figure}

Before the MFM analysis was conducted, the results of the donor simulations were examined.
{In figure \ref{subfig:donor_phi}, mixedness is observed to reach a value of around $0.6$ but appears to not have converged yet. 
Figure \ref{subfig:donor_alpha} shows the two definitions of $\alpha$ over time.
The first definition, $\alpha=h/Agt^2$, reaches a value of about $0.05$ by the end of the simulation, but it does not appear to be converged.
The second definition, $\alpha=\dot{h}^2/4Agh$, is oscillatory, due to the sensitivity of the time derivative to noise, and it appears to fluctuate about a value of approximately $0.04$.
It is acknowledged that this behavior indicates that the RTI simulated here is only weakly turbulent.
However, it is observed that the flow is still self-similar at late times.
The contour plot of $\langle Y_H \rangle $ in figure \ref{fig:YH_contour} exhibits parallel contour lines after $\tau\approx 17$, indicating self-similarity at those times.
It is also shown in the Appendix that the mean concentration and normalized turbulent scalar flux profiles exhibit self-similar collapse after $\tau\approx17$, so the presented self-similar analysis is valid.}

\begin{figure}
    \centering
    \includegraphics[width=0.49\textwidth]{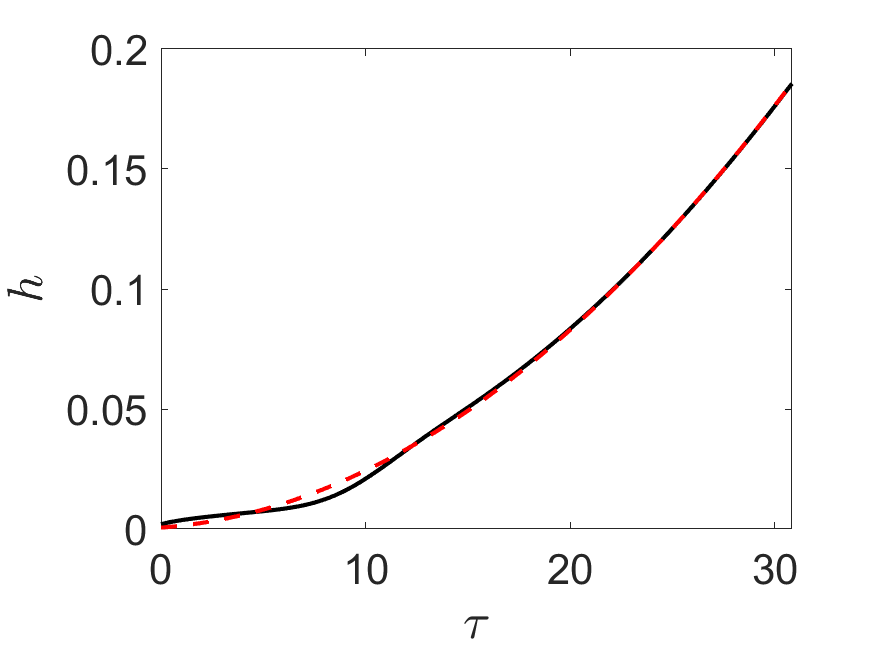}
    \caption{{Black solid line: $h$ measured from DNS; red dashed line: $h_\text{fit}$.}}
    \label{fig:hfit}
\end{figure}
{Figure \ref{fig:hfit} shows a plot of the algebraic fit for $h$, described in equation \ref{eq:hfit}.
For the simulations presented here, $\alpha^*$ is $0.0046$ and $t^*$ is $-1.6\times 10^3$.
The plot shows a strong quadratic dependence of $h$ on $t$ at late time, as $h_\text{fit}$ matches DNS at $\tau \gtrapprox 17$, validating the self-similar ansatz of $h\sim t^2$.}

\begin{figure}
    \centering
    \begin{subfigure}[]{0.49\textwidth}
        \includegraphics[width=\textwidth]{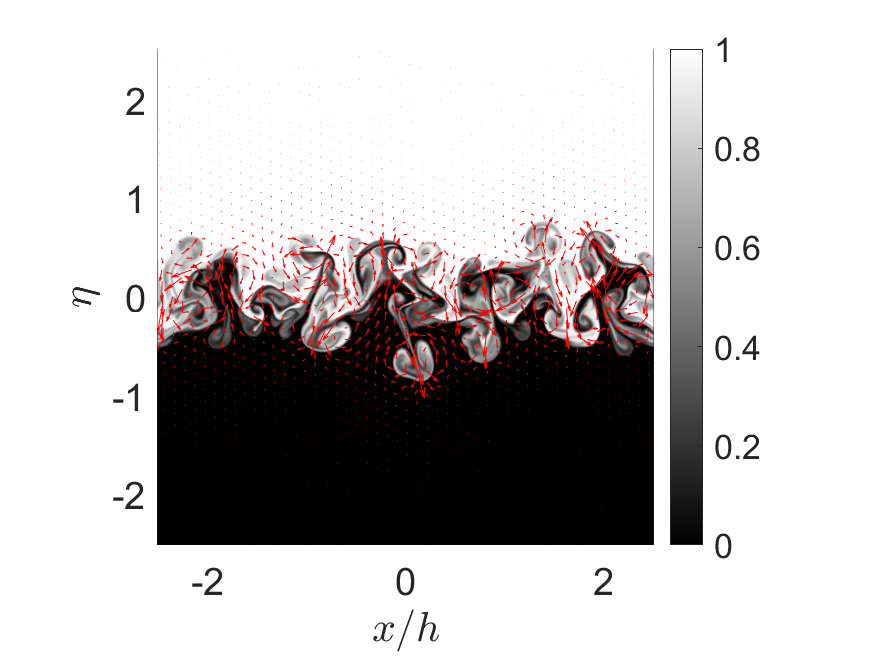}
        \subcaption[]{}
        \label{subfig:donor_field}
    \end{subfigure}
    \begin{subfigure}[]{0.49\textwidth}
        \includegraphics[width=\textwidth]{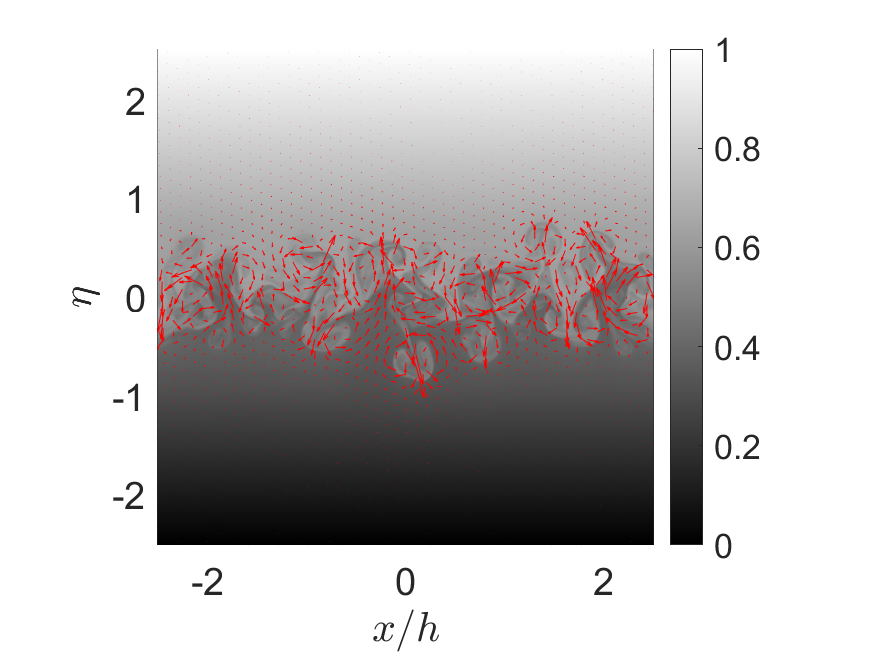}
        \subcaption[]{}
        \label{subfig:receiver_field}
    \end{subfigure}
    \caption{$Y_H$ contours (black: light, white: heavy) and velocity vector fields (red arrows) of the (a) donor simulation and (b) receiver simulation with $s$ enforcing $\langle Y_H \rangle =y$. These snapshots are taken at the last timestep.}
    \label{fig:donorvsreceiver}
\end{figure}

\begin{figure}
    \centering
    \begin{subfigure}[]{0.49\textwidth}
        \includegraphics[width=\textwidth]{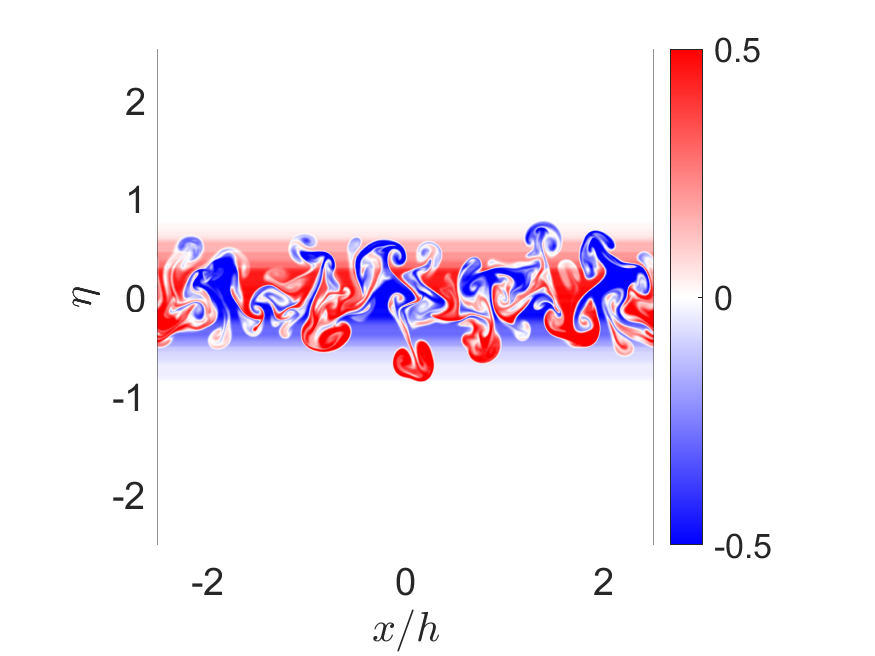}
        \subcaption[]{}
        \label{subfig:donor_field_YHp}
    \end{subfigure}
    \begin{subfigure}[]{0.49\textwidth}
        \includegraphics[width=\textwidth]{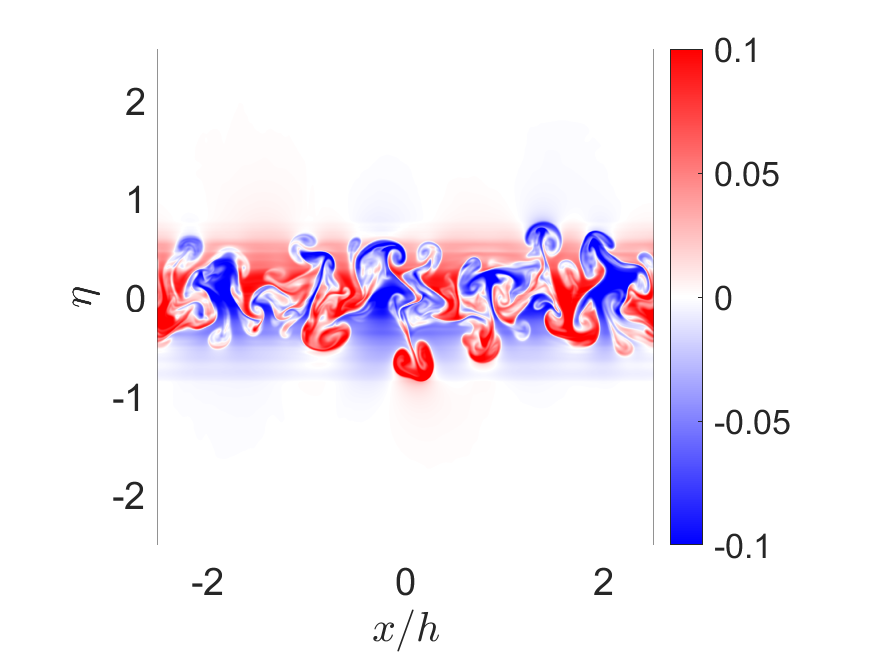}
        \subcaption[]{}
        \label{subfig:receiver_field_YHp}
    \end{subfigure}
    \caption{$Y_H'$ contours of the (a) donor simulation and (b) receiver simulation. These snapshots are taken at the last timestep. {Note that different colorbars have been used to improve interpretability.}}
    \label{fig:donorvsreceiver_YHp}
\end{figure}

To further ensure the simulations are working as desired, the flow fields of the donor and receiver simulations can be examined qualitatively.
The $Y_H$ contours at the last timesteps of each simulation are shown in figure \ref{fig:donorvsreceiver}.
The receiver simulation shown is the one used to compute $D^{00}$ (where $\langle Y_H \rangle =y$).
Self-similar RTI turbulent mixing is observed at this timestep, where the characteristic heavy spikes are sinking into the lighter fluid and the light bubbles rise into the heavier fluid.
Both simulations have the same velocity fields, since the receiver simulation ``receives'' the velocity field from the donor simulation.
In contrast with the donor simulation, which has a stark black-and-white difference between the heavy and light fluids, there is a grey gradient of density in the receiver simulation due to the imposed mean scalar gradient.
The fluctuations of $Y_H$ in each simualtion are also compared in figure \ref{fig:donorvsreceiver_YHp}.
The $Y_H'$ contours are not identical but are qualitatively very similar.
In both simulations, $Y_H'$ is constrained to the mixing layer.
Based on these observations, it is concluded that the simulations are visually working as intended.

\section{Results}
\label{sec:results}
\subsection{Eddy diffusivity moments}
\label{sec:moments}

\begin{figure}
    \centering
    \begin{subfigure}[]{0.49\textwidth}
        \includegraphics[width=\textwidth]{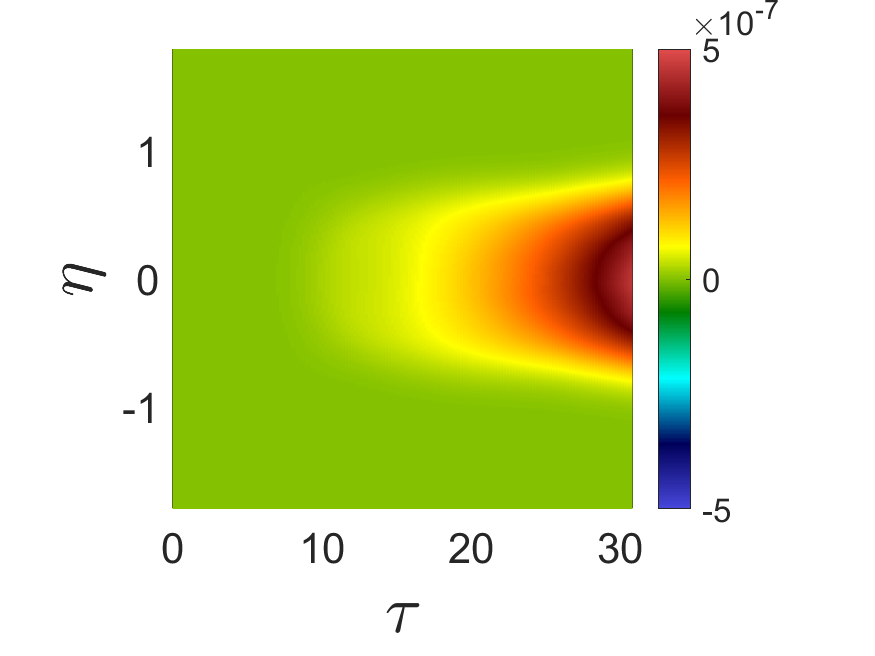}
        \subcaption[]{$D^{00}$}
    \end{subfigure}
    \begin{subfigure}[]{0.49\textwidth}
        \includegraphics[width=\textwidth]{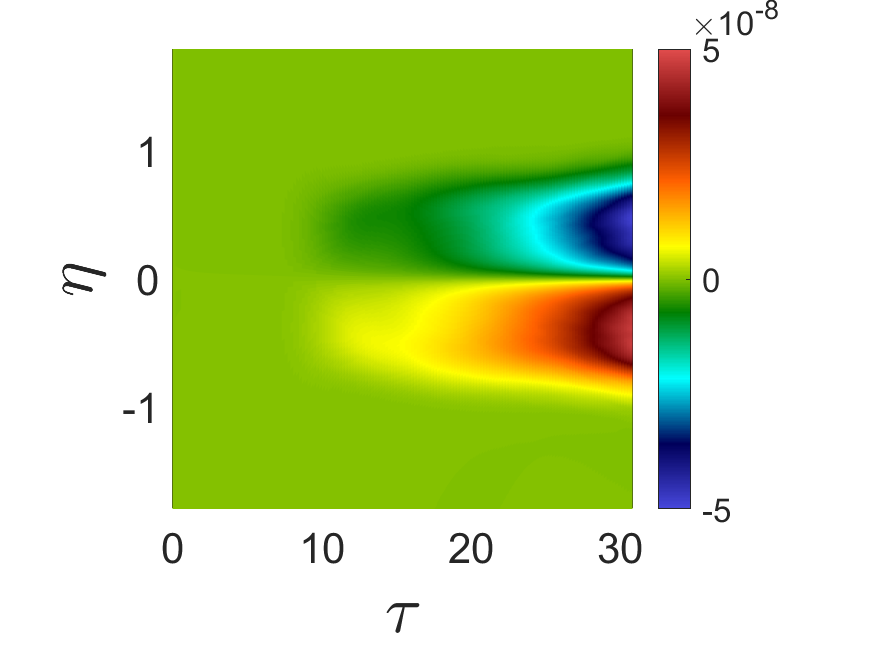}
        \subcaption[]{$D^{10}/h(t)$}
    \end{subfigure}
    \begin{subfigure}[]{0.49\textwidth}
        \includegraphics[width=\textwidth]{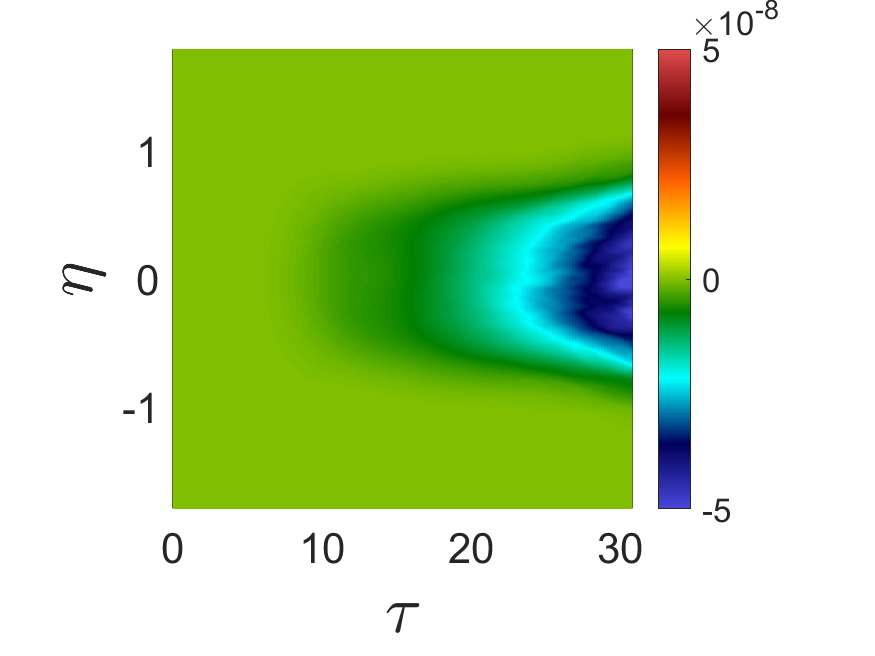}
        \subcaption[]{$D^{01}/t$}
    \end{subfigure}
    \begin{subfigure}[]{0.49\textwidth}
        \includegraphics[width=\textwidth]{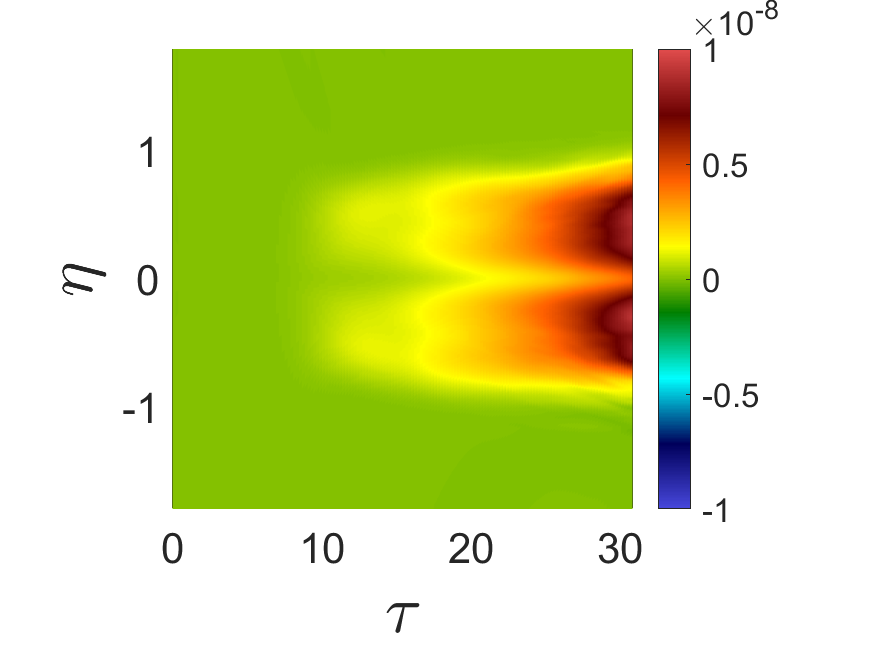}
        \subcaption[]{$D^{20}/h(t)^2$}
    \end{subfigure}
    \caption{Moments of eddy diffusivity kernel normalized by appropriate length and timescales. Data averaged over 1,000 realizations and homogeneous direction $x$.}
    \label{fig:moments}
\end{figure}

Figure \ref{fig:moments} shows normalized MFM measurements of the eddy diffusivity moments $D^{00}$, $D^{10}$, $D^{01}$, and $D^{20}$ averaged over $1,000$ realizations and the homogeneous direction $x$.
Some expected characteristics of the measured moments are observed: 
\begin{enumerate}
    \item {The leading order moment is over two magnitudes larger than the molecular diffusivity ($10^{-9} m^2/s$).
    The scaled higher order moments shown are all at least one magnitude larger than the molecular diffusivity.}
    \item $D^{00}$ is symmetric and always positive.
    \item $D^{10}$ is antisymmetric. 
    {This antisymmetry can be understood by interpreting $D^{10}/D^{00}$ as the centroid of the eddy diffusivity kernel. 
    Physically, for $\eta>0$, it is expected that the mean scalar gradient at the center of the mixing layer (at a negative distance away) has more influence on the turbulent scalar flux than the mean scalar gradient at the outer edges, since the mixing layer is growing outwards.
    This makes the eddy diffusivity kernel skewed more towards the center of the domain, so $D^{10}<0$ for $\eta>0$.
    A similar effect occurs for $\eta<0$, which results in $D^{10}>0$.}
    \item $D^{01}$ is symmetric and always negative. The latter must be true for the flow to depend on its history (it does not violate causality).
    \item $D^{20}$ is symmetric and always positive, as is characteristic of moment of inertia of a positive kernel.
\end{enumerate}

Based on the magnitudes of the normalized moments, some initial observations on importance of each moment can be made.
$D^{00}$ has the highest magnitude of all the moments, which is expected since it is the leading-order moment.
The magnitudes of $D^{10}/h$ and $D^{01}/t$ are on the order of $10\%$ of the magnitude of $D^{00}$, which suggests that they are non-negligible.
On the other hand, the magnitude $D^{20}/h^2$ is on the order of $1\%$ that of $D^{00}$, so $D^{20}$ is likely not an important moment to include in modeling RTI.
More robust studies will be presented in the following sections to determine importance of each of the eddy diffusivity moments.

\begin{figure}
    \centering
    \begin{subfigure}[]{0.49\textwidth}
        \includegraphics[width=\textwidth]{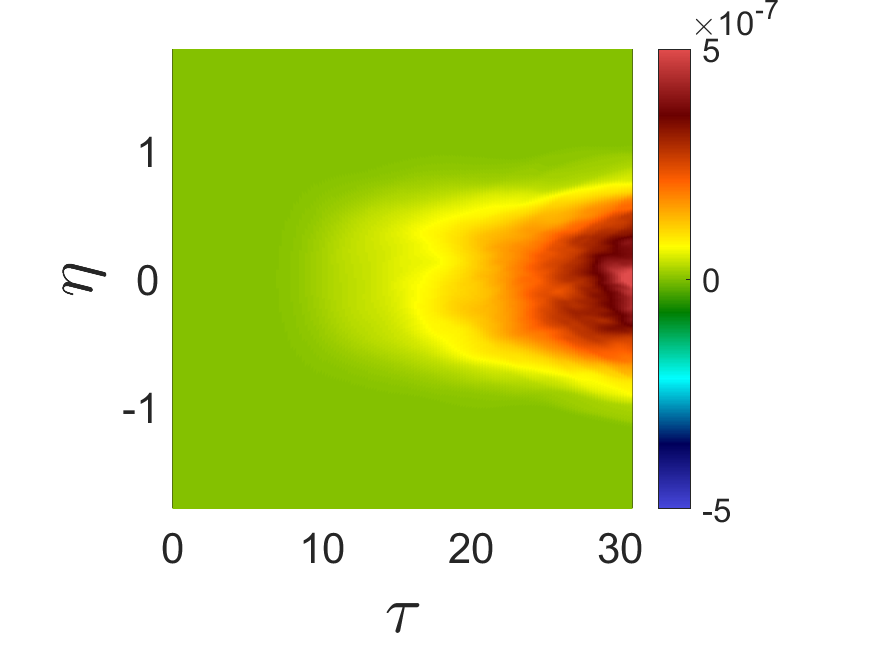}
        \subcaption[]{$8$ realizations}
    \end{subfigure}
    \begin{subfigure}[]{0.49\textwidth}
        \includegraphics[width=\textwidth]{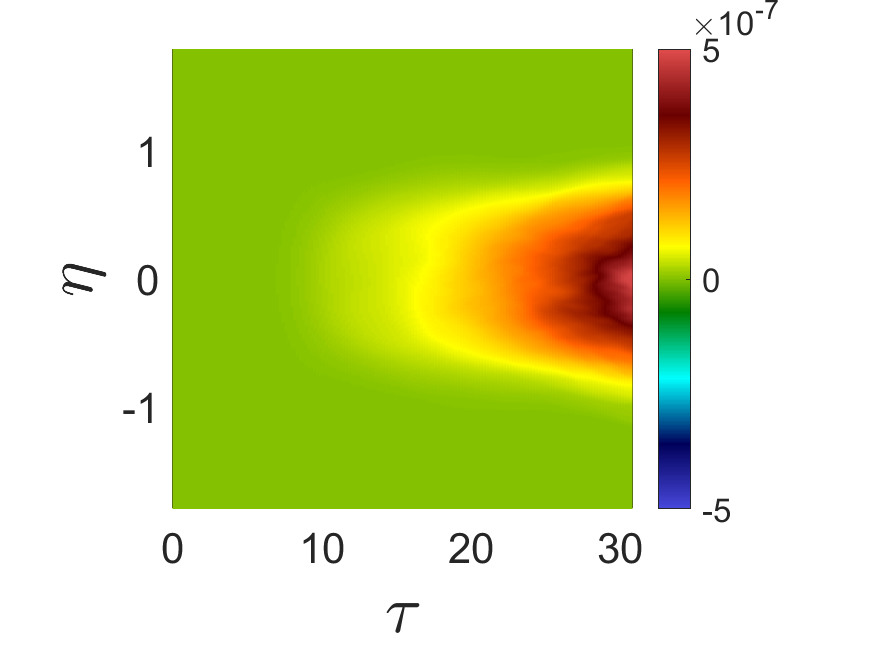}
        \subcaption[]{$32$ realizations}
    \end{subfigure}
    \begin{subfigure}[]{0.49\textwidth}
        \includegraphics[width=\textwidth]{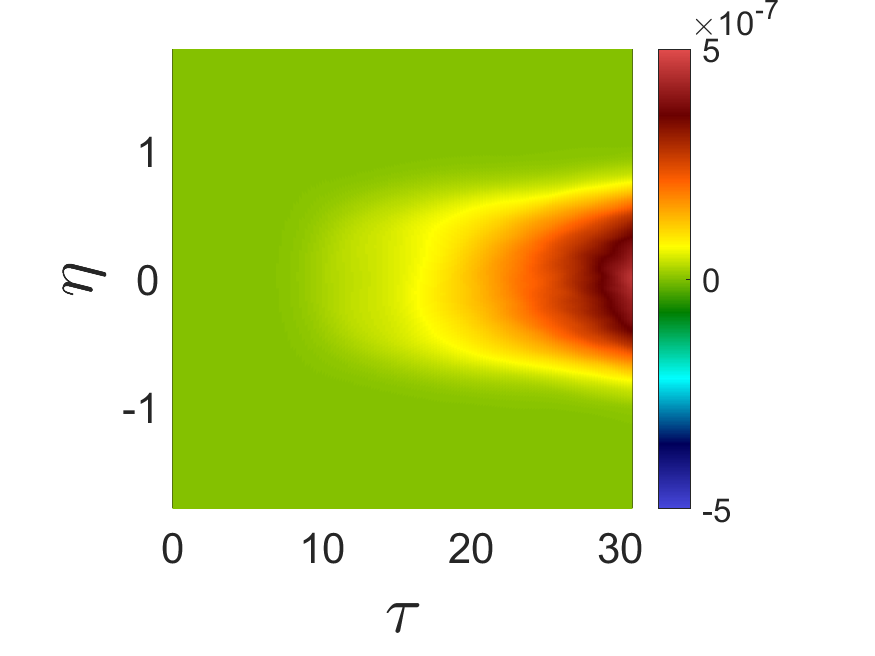}
        \subcaption[]{$100$ realizations}
    \end{subfigure}
    \begin{subfigure}[]{0.49\textwidth}
        \includegraphics[width=\textwidth]{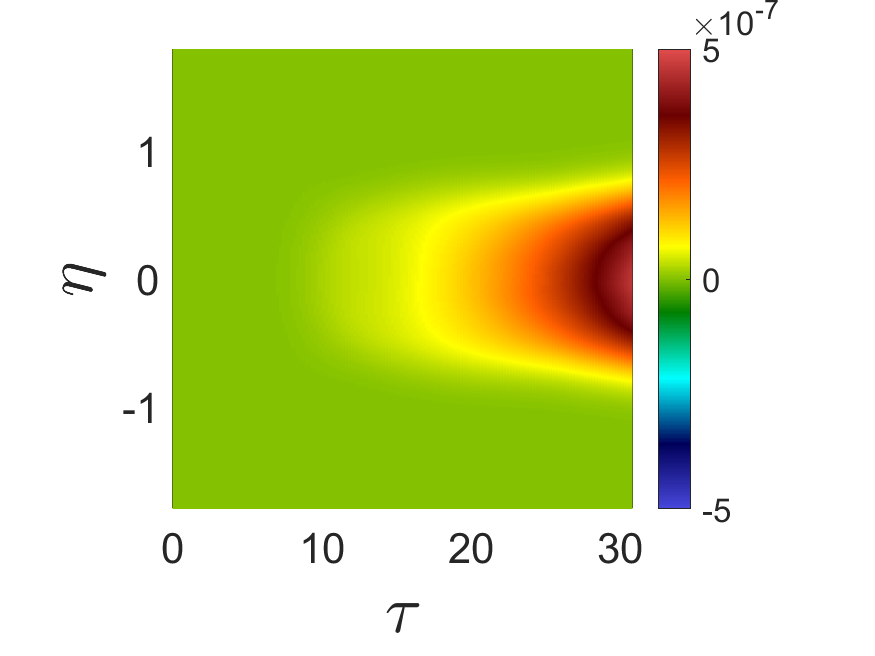}
        \subcaption[]{$1000$ realizations}
    \end{subfigure}
    \caption{$D^{00}$ averaged over different numbers of realizations.}
    \label{fig:stat_conv}
\end{figure}

It is also observed that there is statistical error in the measurements.
Due the chaotic nature of RTI, the moment measurements contain statistical error, but this error can be reduced by averaging many realizations.
To demonstrate statistical convergence of the measurements, plots of $D^{00}$ averaged over different numbers of realizations are included in figure \ref{fig:stat_conv}.
As the number of realizations increases, the plots become smoother, and it is found that after $1,000$ realizations, the rate of reduction in statistical error slows down significantly.
Averaging over this number of realizations results in a smooth $D^{00}$ measurement and higher-order moment measurements with acceptably less statistical error.

{Additionally,} the higher the order of the moment, the slower its rate of statistical convergence. 
Recall that determination of higher-order moments requires information from lower-order moments.
For example, in determining $D^{01}$, $tD^{00}$ is subtracted from $F^{01}$, the turbulent scalar flux measurement in the simulation associated with $D^{01}$.
Naturally, there is statistical error associated with both $D^{01}$ and $D^{00}$.
However, the error in $D^{00}$ is amplified by $t$, so the overall statistical error of $D^{01}$ increases with time.
This statistical error ``leakage'' occurs for all higher-order moments.
The higher the order of the moment, the worse the statistical error, since information from more lower-order moments is needed, and so more statistical error is accumulated and amplified.
The relatively high statistical error of the higher-order moments makes it challenging to study their importance.
Particularly, taking derivatives of quantities with high statistical error amplifies the error, so smoother measurements are desired.
In this work, the moment measurements are smoothed using a Savitzky-Golay filter function in Matlab with a polynomial order of unity and window size of $191$.
These smoothed moments are shown in figure \ref{fig:sm_moments}.
While it is possible to design an alternative formulation of MFM that removes leakage of statistical error from low-order moments to higher-order moments \citep[see][]{lavacot2022}, for this 2D study and for the order of moments considered here, the statistical convergence is sufficient. 

\begin{figure}
    \centering
    \begin{subfigure}[]{0.49\textwidth}
        \includegraphics[width=\textwidth]{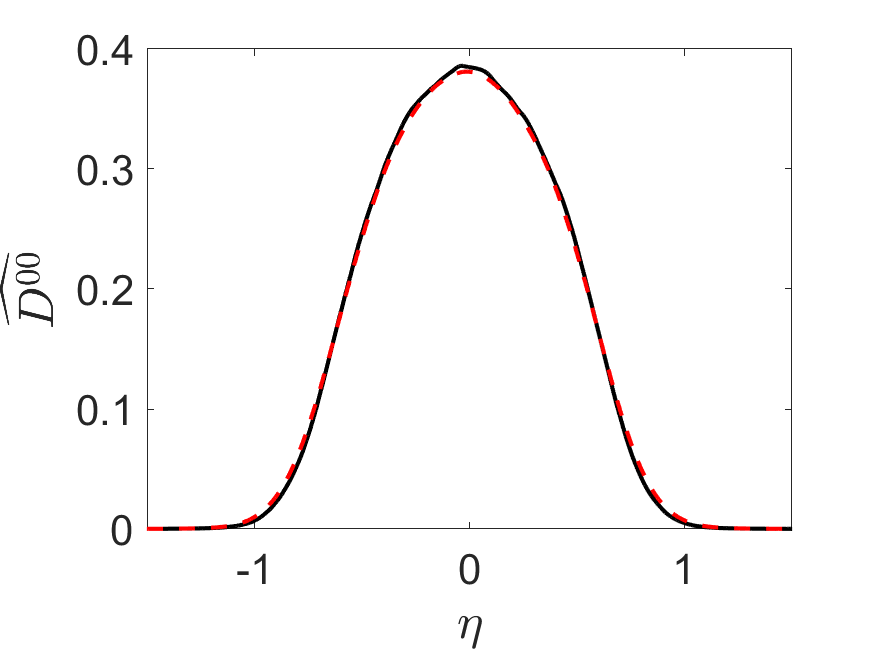}
        \subcaption[]{}
    \end{subfigure}
    \begin{subfigure}[]{0.49\textwidth}
        \includegraphics[width=\textwidth]{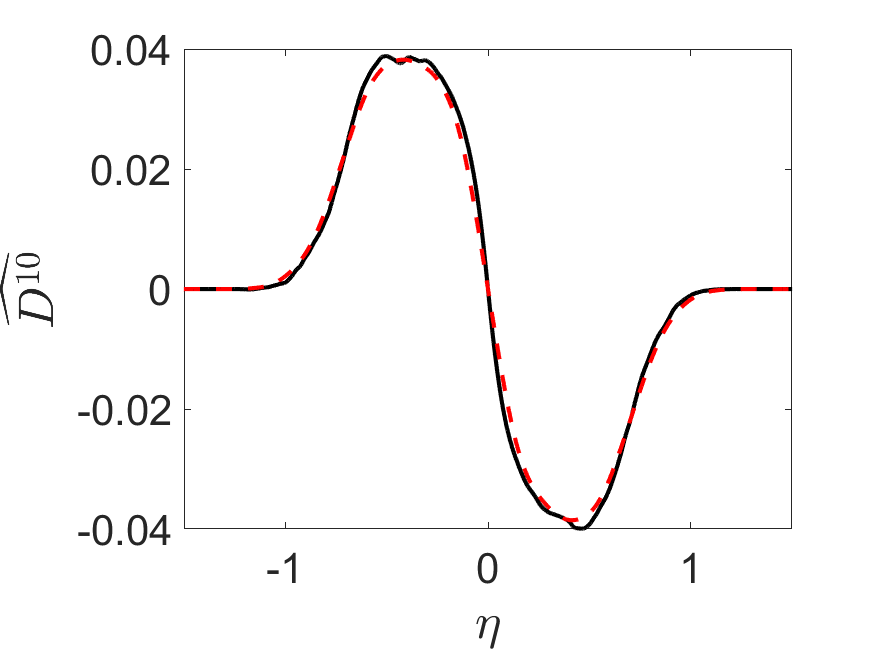}
        \subcaption[]{}
    \end{subfigure}
    \begin{subfigure}[]{0.49\textwidth}
        \includegraphics[width=\textwidth]{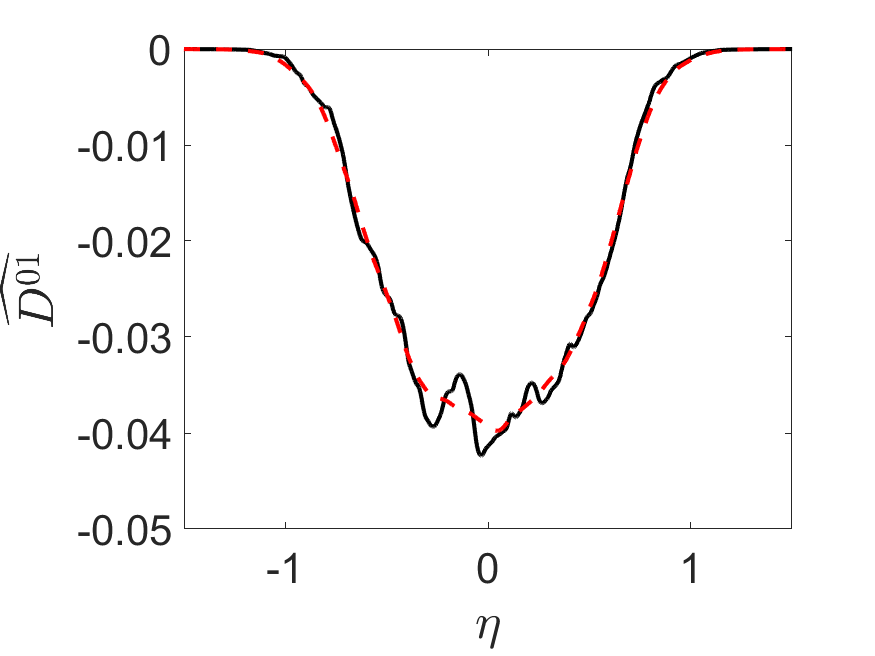}
        \subcaption[]{}
    \end{subfigure}
    \begin{subfigure}[]{0.49\textwidth}
        \includegraphics[width=\textwidth]{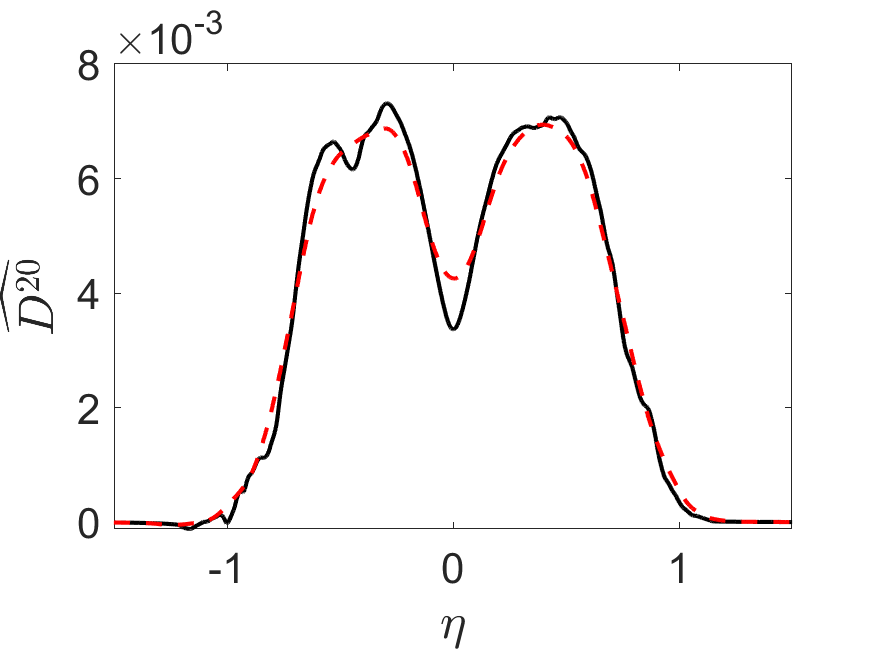}
        \subcaption[]{}
    \end{subfigure}
    \caption{Smoothed moments (dashed red) over raw MFM measurements of moments (solid black). The moments are taken from the mean data at the last timestep of the simulations and are transformed to self-similar space.}
    \label{fig:sm_moments}
\end{figure}

{Using these measurements, nonlocal timescales and lengthscales ($t_{NL}$ and $L_{NL}$, respectively) can be defined:
\begin{align}
    t_{NL} = -\frac{D^{01}}{D^{00}}, \quad L_{NL} = \sqrt{\frac{D^{20}}{D^{00}}}.
\end{align}
Note that this analysis can only be done for $-1\leq\eta\leq 1$, since the moments are analytically zero outside the mixing layer.}

\begin{figure}
    \centering
    \begin{subfigure}[]{0.49\textwidth}
        \includegraphics[width=\textwidth]{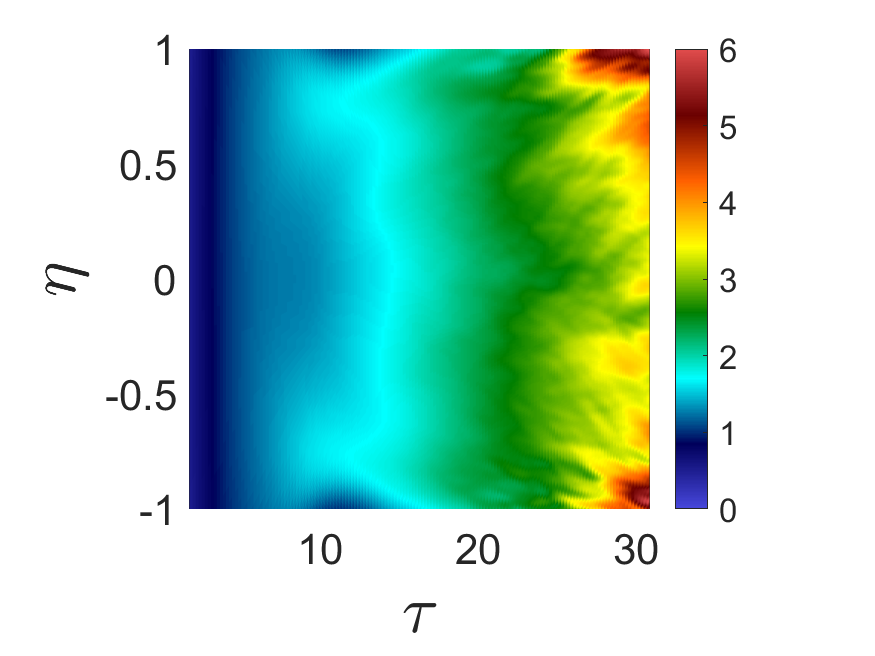}
        \subcaption[]{$\tau_{NL}$}
    \end{subfigure}
    \begin{subfigure}[]{0.49\textwidth}
        \includegraphics[width=\textwidth]{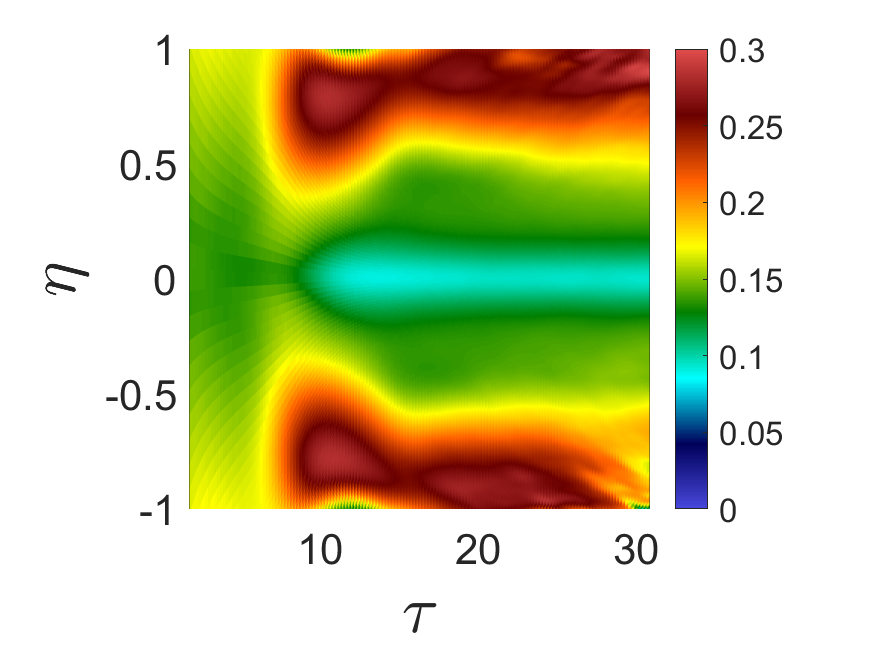}
        \subcaption[]{$\eta_{NL}$}
    \end{subfigure}
    \caption{{Nondimensional nonlocal timescale and lengthscale contours. Only $-1\leq\eta\leq 1$ is plotted, since moments are zero outside the mixing layer. Early times ($\tau<2$) not plotted due to transient behavior. }}
    \label{fig:nonlocal_contours}
\end{figure}

\begin{figure}
    \centering
    \begin{subfigure}[]{0.49\textwidth}
        \includegraphics[width=\textwidth]{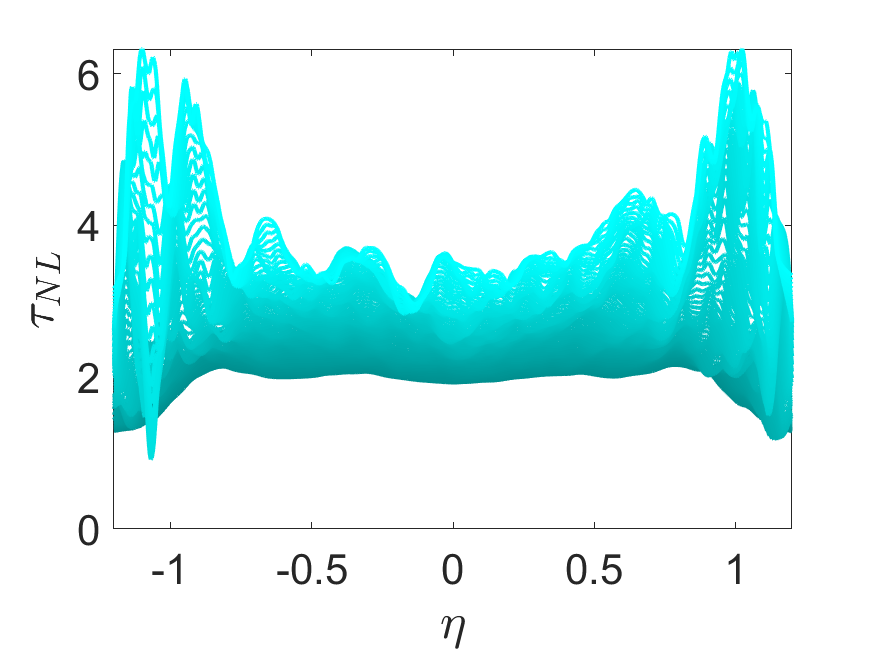}
        \subcaption[]{}
    \end{subfigure}
    \begin{subfigure}[]{0.49\textwidth}
        \includegraphics[width=\textwidth]{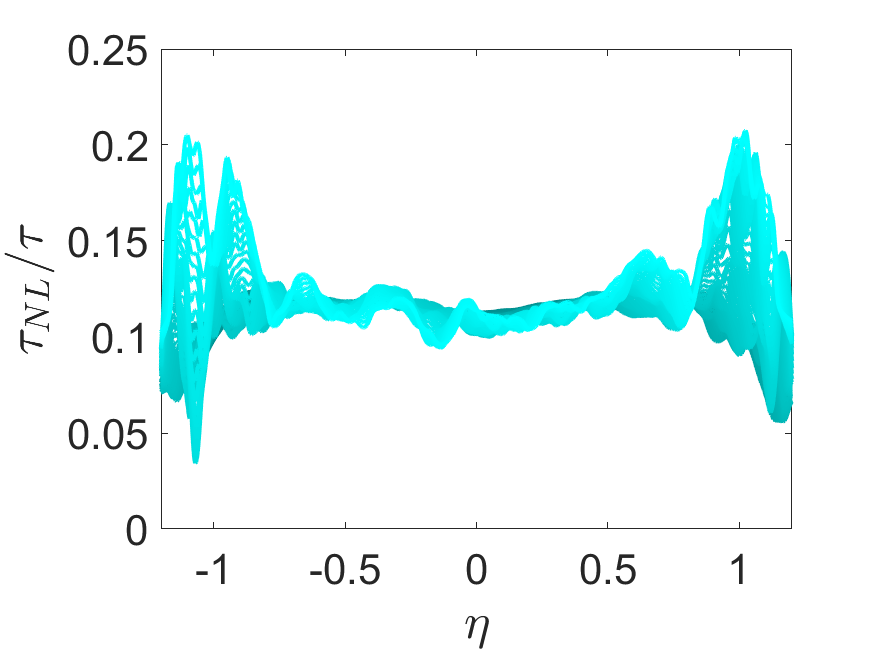}
        \subcaption[]{}
    \end{subfigure}
    \caption{{Nondimensional nonlocal timescale profiles at different times for $\tau>17$. Lighter lines correspond to later times; darker lines correspond to earlier times. (a) is unscaled and shows the linear time dependence of $\tau_{NL}$ on $\tau$. (b) shows the collapse of the profiles when scaled by $\tau$.}}
    \label{fig:nonlocal_timescale}
\end{figure}

\begin{figure}
    \centering
    \includegraphics[width=0.49\textwidth]{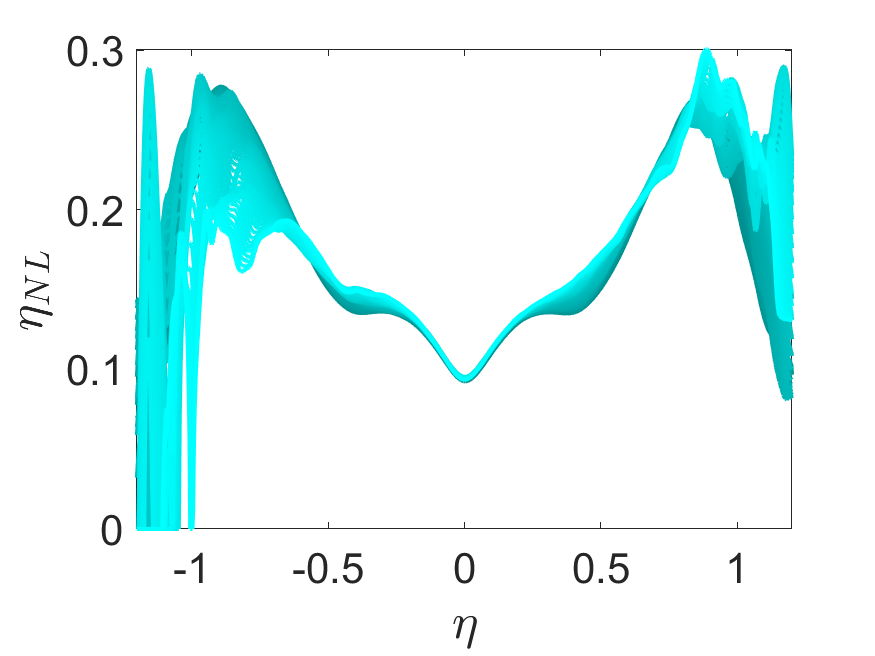}
    \caption{{Nondimensional nonlocal lengthscale profiles at different times for $\tau>17$. Lighter lines correspond to later times; darker lines correspond to earlier times. }}
    \label{fig:nonlocal_lengthscale}
\end{figure}

{Nondimensionally, the nonlocal timescale is $\tau_{NL} = t_{NL}/t_0$, and the nonlocal lengthscale is $\eta_{NL}=L_{NL}/h$.
Contour plots of the nondimensionalized nonlocal timescale and lengthscale are in figure 
\ref{fig:nonlocal_contours}.
Note that $\tau_{NL}$ scales as $\tau$, so profiles of $\tau_{NL}/\tau$ against $\eta$ are also plotted in figure \ref{fig:nonlocal_timescale} in the self-similar time regime ($\tau>17$).
The scaled profiles collapse and have a centerline value of approcimately $0.1$.
This means that the mean fluxes at some time $\tau$ are affected by mean scalar gradients $0.1\tau$ earlier.
Figure \ref{fig:nonlocal_lengthscale} shows the minimum nonlocal lengthscale is at the centerline, where $\eta_{NL}\approx0.09$. 
The maximum lengthscales occur near the outer edges of the mixing layer: at around $\eta=\pm 0.87$, $\eta_{NL}\approx0.27$.
This indicates that mean fluxes at the mixing layer edges depend mostly on mean scalar gradients about a quarter of a mixing width away, while mean fluxes at the centerline depend on mean scalar gradients about one tenth of a mixing width away; nonlocality appears to be stronger at the mixing layer edges than at the centerline.
These nonlocal properties of the eddy diffusivity for RTI could not be predicted without direct measurement of the eddy diffusivity moments, which has been made possible through MFM.}

\subsection{Assessment of importance of nonlocal effects}

\subsubsection{Comparison of terms in turbulent scalar flux expansion}
\label{sec:term_comp}

\begin{figure}
    \centering
    \includegraphics[width=0.45\textwidth]{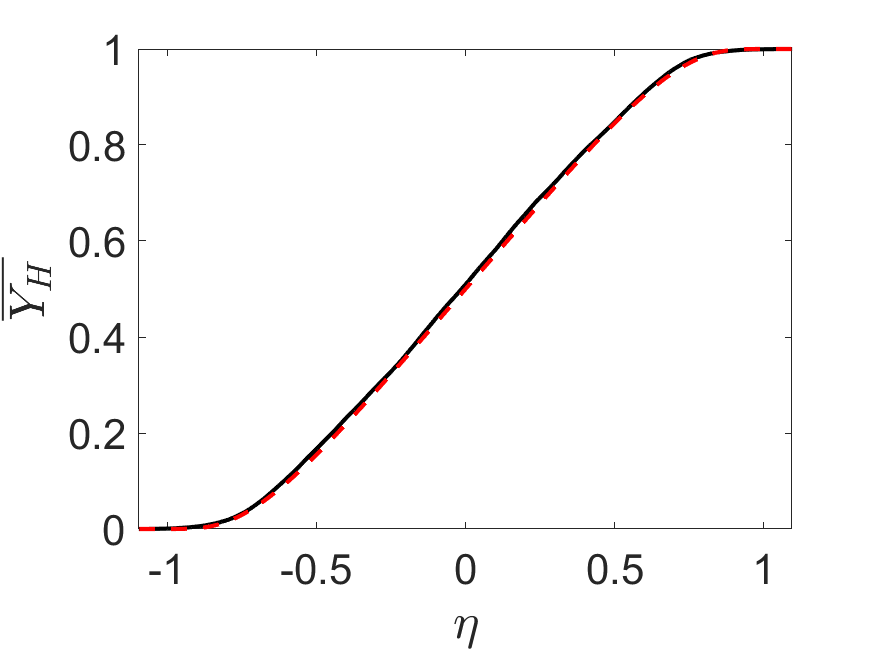}
    \caption{Semi-analytical fit to $\langle Y_H \rangle $ (dashed red) against DNS measurement of $\langle Y_H \rangle $ (solid black).}
    \label{fig:YH_fit}
\end{figure}

To aid in the determination of which moments are important for a RANS model, a comparison of the terms in the expansion of the turbulent scalar flux (equation \ref{eq:explicit}) is presented.
These terms involve gradients of $\langle Y_H \rangle $.
Instead of using $\langle Y_H \rangle $ directly from the DNS, a fit to $\langle Y_H \rangle $  is used, since the statistical error in the raw measurement gets amplified by derivatives in $\eta$.
That is, the quantities of interest are sufficiently converged for plotting but not for operations involving derivatives.
Thus, an analytical fit to $\langle Y_H \rangle $ is obtained as follows:
\begin{gather}
    \langle Y_H \rangle ^* = \begin{cases}
    0 & \text{if } \eta<-a\\
    \int_{-a}^\eta \frac{1}{\left(a^2-{\eta'}^2\right)^2}\exp\left(\frac{1}{B\left(a^2-{\eta'}^2\right)}\right)d{\eta'} & \text{if } -a\leq\eta\leq a\\
    1 & \text{if } \eta>a
    \end{cases},\\
    \langle Y_H \rangle  = \frac{\langle Y_H \rangle ^*}{\langle Y_H \rangle ^*_\text{max}},
\end{gather}
where the integral is determined numerically, and $a$ and $B$ are fitting coefficients.
The coefficients $a^2=1.2$ and $B=0.36$ are found to give good agreement to the mean concentration profile from DNS, as shown in figure \ref{fig:YH_fit}.

\begin{figure}
    \centering
    \includegraphics[width=0.5\textwidth]{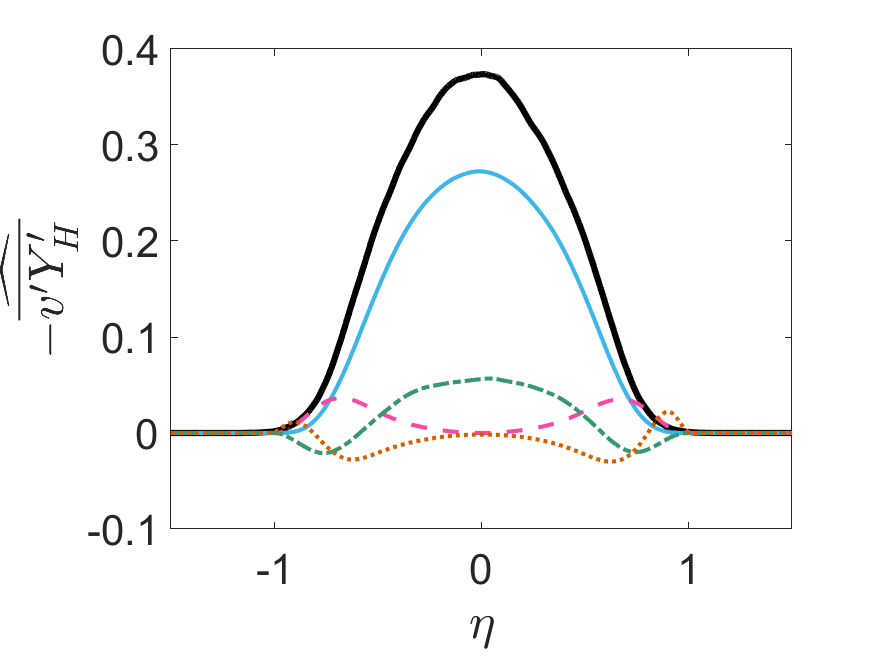}
    \caption{Comparison of terms in the expansion of the turbulent scalar flux. Black: DNS measurement of turbulent scalar flux, solid blue: $\widehat{D^{00}}$ term, dashed pink: $\widehat{D^{10}}$ term, dash-dotted green: $\widehat{D^{01}}$ term, dotted orange: $\widehat{D^{20}}$ term.}
    \label{fig:exp_terms}
\end{figure}
The terms on the right hand side of equation \ref{eq:explicit} are plotted against the DNS measurement of the turbulent scalar flux in figure \ref{fig:exp_terms}.
Clearly, the $\widehat{D^{00}}$ term is not enough to capture the turbulent scalar flux.
It is observed that the $\widehat{D^{01}}$ term is significant in magnitude in the middle of the domain, and {the $\widehat{D^{10}}$ term carries importance at the outer edges of the mixing layer.
The term associated with the highest-order moment that was measured, $\widehat{D^{20}}$ also appears to be of similar magnitude as the other moments, indicating it may also carry important information about nonlocality of the eddy diffusivity.
These preliminary findings indicate nonlocality is certainly important for accurate modeling of mean scalar transport in this RTI problem, since the higher-order terms in equation \ref{eq:explicit} appear non-negligible compared to the leading-order term.
It may be tempting to ascribe physical reasons for the behavior of the terms plotted in figure \ref{fig:exp_terms}, but this is not so straightforward, especially since the full eddy diffusivity kernel for this problem as not yet been measured.
Further, it would be inappropriate to draw conclusions about importance of each eddy diffusivity moment in a RANS model, since the operator form must be scrutinized first.
A faulty operator form could give misleading implications about certain eddy diffusivity moments.
It turns out that a simple superposition of these terms, which would represent a truncation of equation \ref{eq:explicit}, does not accurately represent the true eddy diffusivity kernel and actually leads to divergence of predictions, so such an operator form would not be appropriate; this will be covered more in depth later in \S \ref{sec:improve_RANS}. 
Nevertheless, the results shown here are strong evidence of nonlocality of the eddy diffusivity kernel for the RTI simulated here.}

\subsubsection{Comparison of leading-order model against a local model}
\begin{figure}
    \centering
    \begin{subfigure}[]{0.49\textwidth}
        \includegraphics[width=\textwidth]{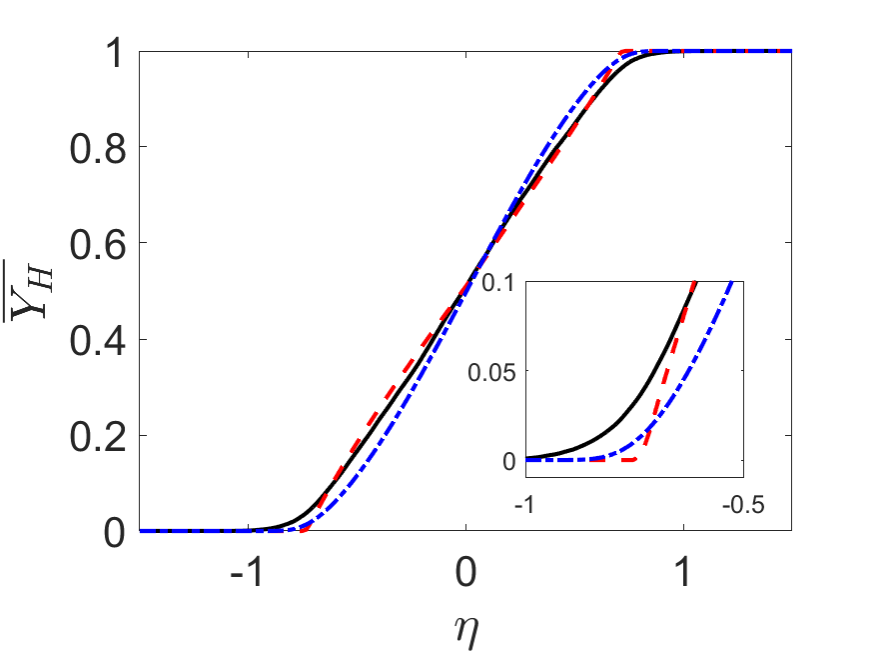}
        \subcaption[]{}
        \label{fig:klvsmfm_YH}
    \end{subfigure}
    \begin{subfigure}[]{0.49\textwidth}
        \includegraphics[width=\textwidth]{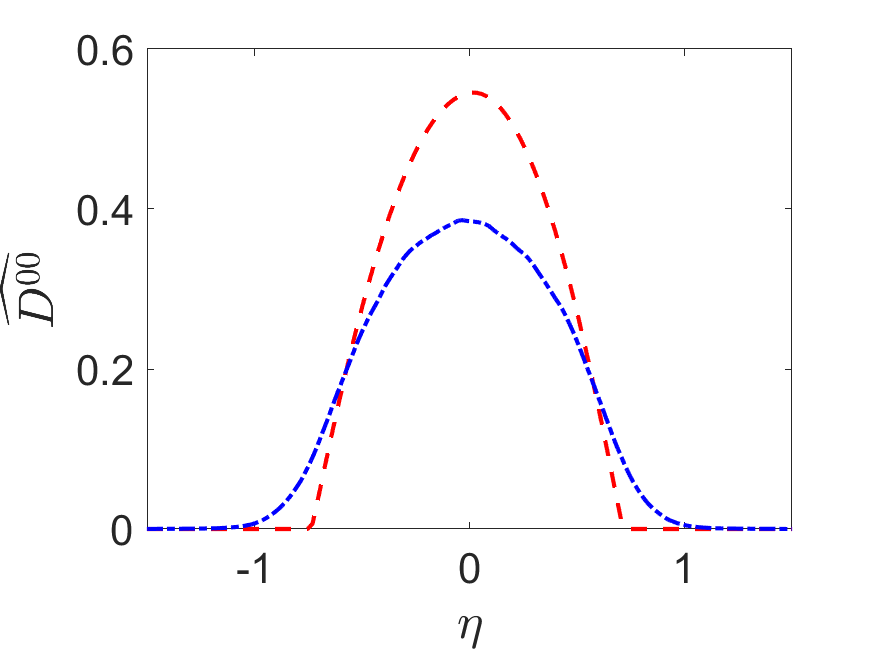}
        \subcaption[]{}
        \label{fig:klvsmfm_D00}
    \end{subfigure}
    \caption{Comparisons of (a) model coefficient from $k$-$L$ model with leading-order moment measured using MFM and (b) resulting similarity solutions for $\langle Y_H \rangle $.
    Inset plot in (a) shows zoomed-in view around $\eta=-1$ to highlight divergence of $k$-$L$ model from DNS.
    Solid black: DNS measurement, dash-dotted blue: leading-order MFM-based model, dashed red: $k\text{-}L$ model.}
    \label{fig:klvsmfm}
\end{figure}

To demonstrate the shortcomings of models using purely local coefficients, an MFM-based leading-order model and the $k$-$L$ RANS model are compared.
{The intent of this study is not to immediately propose a ``better'' RANS model to replace $k$-$L$, nor is it to suggest the MFM-based leading-order model is more accurate than the $k$-$L$ model.
In fact, it is expected that the MFM-based leading-order model will perform poorly, since it does not include important higher-order moments of eddy diffusivity.
Instead, this study emphasizes the necessity of higher-order moments and shows how MFM can reveal  incorrect model forms.}

In particular, a 1D $k$-$L$ simulation is run, and the eddy diffusivity and mean concentration profiles are extracted from the results to be compared to those of the MFM-based model using the measured $D^{00}$ that was presented in \S \ref{sec:moments}.
The $k$-$L$ simulation used in this section is implemented in Ares, and details of the implementation are in \citet{morgangreenough2015} and \citet{morgan2018response}.
Note that the $k$-$L$ simulation is used here for illustration purposes and should not be confused with the 2D DNS simulations used to obtain our MFM moments.

The MFM-measured $D^{00}$ is used for the leading-order MFM-based model:
\begin{align}
    -\langle v'Y_H' \rangle  = D^{00}_\text{MFM}\frac{\partial\langle  Y_H \rangle }{\partial y}.
    \label{eq:MFM_leading_order}
\end{align}
To solve this, $D^{00}_\text{MFM}$ is obtained from the smoothed MFM measurements and transformed to self-similar coordinates.
The resulting $\widehat{D^{00}_\text{MFM}}$ is a function of $\eta=\frac{y}{h_\text{DNS}}$, where $h_\text{DNS}=\alpha^*_\text{DNS}Ag(t-t^*_\text{DNS})^2$ is an algebraic fit to the mixing width from the DNS.
The equation is then solved semi-analytically in conjunction with the mean mass fraction evolution equation in self-similar coordinates:
\begin{align}
    -2\eta\frac{d\langle Y_H \rangle }{d\eta}=\frac{d}{d\eta}\left(-\widehat{\langle v'Y_H' \rangle }\right),\\
    -\widehat{\langle v'Y_H' \rangle } = \widehat{D^{00}_\text{MFM}}\frac{d\langle Y_H \rangle }{d\eta}\label{eq:MFM_leading_order_ss}.
\end{align}

The $k\text{-}L$ model uses the gradient diffusion approximation for the turbulent flux:
\begin{align}
    -\langle v'Y_H' \rangle  = \frac{\mu_t}{\langle \rho \rangle N_Y}\frac{\partial\langle  Y_H \rangle }{\partial y}\label{eq:kL_grad_diff} = D^{00}_{k\text{-}L}\frac{\partial\langle  Y_H \rangle }{\partial y},
\end{align}
where $\mu_t=C_\mu\langle \rho \rangle L\sqrt{2k}$.
$N_Y$ {is one of the model coefficients} set by similarity constraints derived by \citet{dimontetipton2006}.
{Particularly, this work uses the coefficient calibration detailed in \citep{morgangreenough2015}, and the coefficients are chosen to achieve the same $\alpha$ as the DNS.
Here, $C_\mu$ is unity and $N_Y$ is $2.47$.
The $k$-$L$ RANS model is solved in spatio-temporal coordinates, and the $\langle \rho \rangle $, $k$, and $L$ obtained from the solution are used to compute $\mu_t$ and, consequently, $D^{00}_{k\text{-}L}$, which is purely local.}
For a meaningful comparison with the MFM-based model, $D^{00}_{k\text{-}L}$ is transformed to $\widehat{D^{00}_{k\text{-}L}}$ according to the self-similar coordinate $\xi=\frac{y}{h_{k\text{-}L}}$, where $h_{k\text{-}L}=\alpha^*_{k\text{-}L}Ag(t-t^*_{k\text{-}L})^2$. 
It must be noted that the $h$ fitting coefficients $\alpha^*$ and $t^*$ are not the same between the DNS and $k$-$L$ solutions.
In this work, $\alpha^*_\text{DNS}=0.046$, $t^*_\text{DNS}=-1600\ s$, $\alpha^*_{k\text{-}L}=0.04$, and $t^*_{k\text{-}L}=1250\ s$ ($t^*_{k\text{-}L}$ is positive due to the relaxation time to the self-similar profiles in the beginning of the $k$-$L$ simulation). 

{Figure \ref{fig:klvsmfm_YH} shows the mean concentration profiles computed using each of the two models.
As expected, the MFM-based leading-order model performs poorly, not capturing the slope of the DNS profile, since that model only uses the leading-order eddy diffusivity moment and incorporates no information about nonlocality of the eddy diffusivity.
The $k$-$L$ model exhibits divergence from DNS at the outer edges of the mixing layer, since it is designed to predict a linear $\langle Y_H \rangle$ profile.
However, it does capture the slope of the DNS profile, despite it also using a leading-order closure.
In addition, it is observed in figure \ref{fig:klvsmfm_D00} that the MFM-measured $\widehat{D^{00}_\text{MFM}}$ is significantly lower in magnitude than  $\widehat{D^{00}_{k\text{-}L}}$.
Here, MFM reveals that the $k$-$L$ model is using an incorrect model form, since the $D^{00}$ it is using does not match the MFM measurement.
In fact, the $k\text{-}L$ model is using this higher-magnitude coefficient in order to compensate for the error in model form and achieve a linear mean concentration profile with a slope that matches DNS.
Despite this compensation, the $k$-$L$ model still disagrees with DNS results at the outer edges of the mixing layer, which are important to capturing the average reaction rate in reacting flows like in ICF.
A more accurate RANS model would more closely match the eddy diffusivity moments measured by MFM.
As results will show shortly, the gap between the leading-order MFM-based model $\langle Y_H \rangle $ and the DNS measurement would be bridged by inclusion of higher-order moments, which would introduce information about the nonlocality of the eddy diffusivity.}

\subsection{Assessment of nonlocal {operator} forms}
\label{sec:improve_RANS}
In this section, two RANS {operator} forms using information about the nonlocality of the eddy diffusivity are presented.
These are the explicit and implicit {operator} forms; the former is a truncation of the turbulent scalar flux expansion \ref{eq:explicit}, and the latter will be presented shortly. 
It must be stressed that the intention of the following studies is not to propose a new RANS model.
Ultimately, a RANS model should not depend on direct MFM measurements that can only be retrieved from impractically many DNS.
Instead, these studies are performed to further assess the importance of each of the eddy diffusivity moments, determine which combinations of moments best enhance the performance of a RANS model, and examine the differences between the explicit and implicit {operator} forms.
The aim of these studies is to inform development of more predictive RANS models for RTI, not to suggest that these are the exact models that should be used.

In addition, $D^{20}$ will not be included in the following studies.
This is mainly due to the high statistical error in the measurement that makes it difficult to ascertain whether errors in the results are due to this statistical error or solely the addition of the moment to the model.
From the comparison of terms in \S \ref{sec:term_comp}, it is expected that $D^{20}$ is not as important as $D^{10}$ and $D^{01}$ to include in a RANS model.
This should be tested in future work when a more statistically-converged measurement is achieved for $D^{20}$, ideally in a 3D analysis.

\subsubsection{Explicit {operator} form}
\label{sec:exp_model_form}

The explicit {operator} form is a truncation of the expansion of the turbulent scalar flux, as defined in equation \ref{eq:explicit}.
\citet{hamba1995} and \citet{hamba_2004} have examined this form in the context of shear flows.
Transformation of this expansion to self-similar coordinates and substitution into \ref{eq:ste_avg} results in
\begin{align}
    {-}2\eta \frac{d\langle Y_H \rangle }{d\eta}=\frac{d}{d\eta}\left[\left(\widehat{D^{00}}{-}\widehat{D^{01}}\right)\frac{d\langle Y_H \rangle }{d\eta} + \left({-}\eta\widehat{D^{01}}+\widehat{D^{10}}\right)\frac{d^2\langle Y_H \rangle }{d\eta^2}
    {+\widehat{D^{20}}\frac{d^3\langle Y_H \rangle }{d\eta^3}+\dots}\right],
    \label{eq:explicit_model}
\end{align}
which can be solved numerically for $\langle Y_H \rangle $.
The $\widehat{D^{mn}}$ used in the numerical solve are the smoothed, normalized moments.
To determine which eddy diffusivity moments are important in constructing RANS models for RTI, different combinations of $\widehat{D^{mn}}$ terms are kept in equation \ref{eq:explicit_model}, and the results are compared to DNS.
In the numerical solve, equation \ref{eq:explicit_model} is discretized on a staggered mesh, and derivatives are computed using central finite differences. 
A matrix-vector equation is assembled and solved for $\langle Y_H \rangle $ with Dirichlet boundary conditions.

\begin{figure}
    \centering
    \begin{subfigure}[]{0.49\textwidth}
        \includegraphics[width=\textwidth]{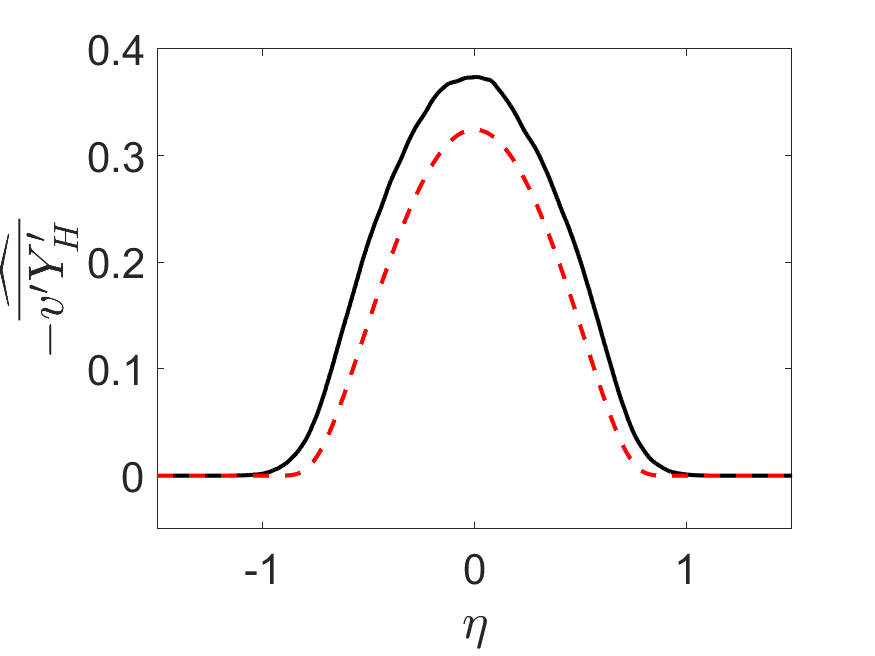}
        \subcaption[]{$D^{00}$}
    \end{subfigure}
    \begin{subfigure}[]{0.49\textwidth}
        \includegraphics[width=\textwidth]{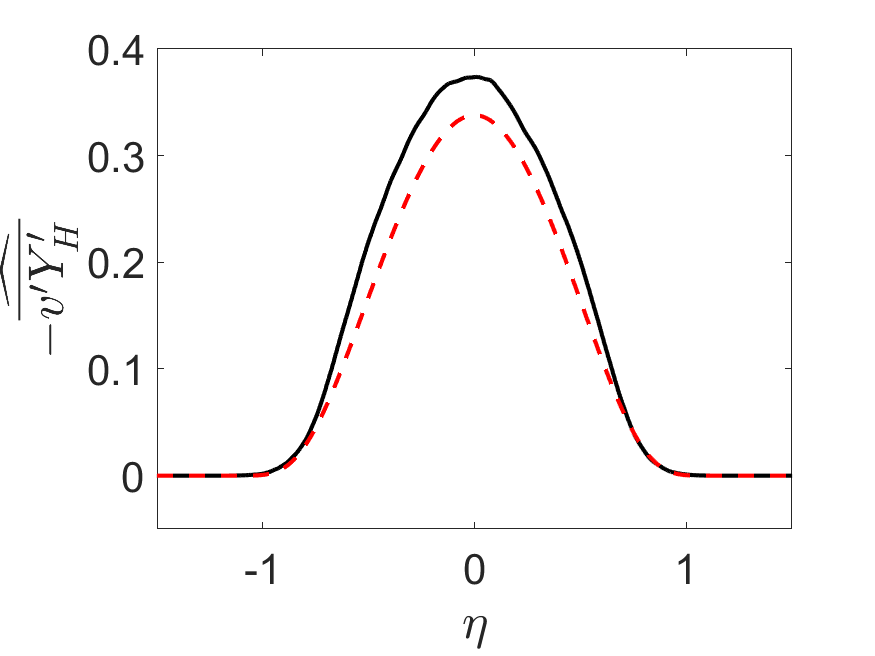}
        \subcaption[]{$D^{00}, D^{10}$}
    \end{subfigure}
    \begin{subfigure}[]{0.49\textwidth}
        \includegraphics[width=\textwidth]{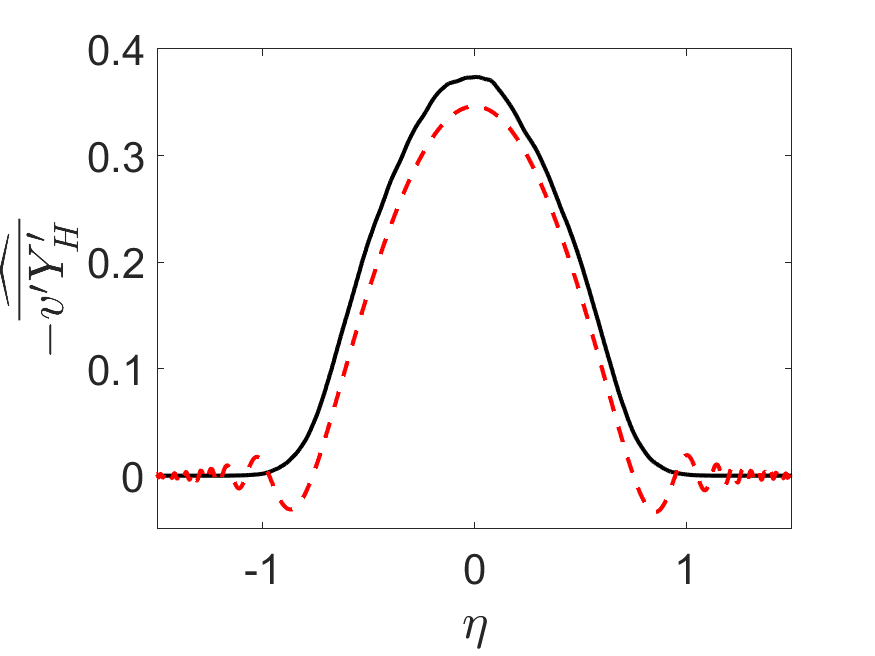}
        \subcaption[]{$D^{00}, D^{01}$}
    \end{subfigure}
    \begin{subfigure}[]{0.49\textwidth}
        \includegraphics[width=\textwidth]{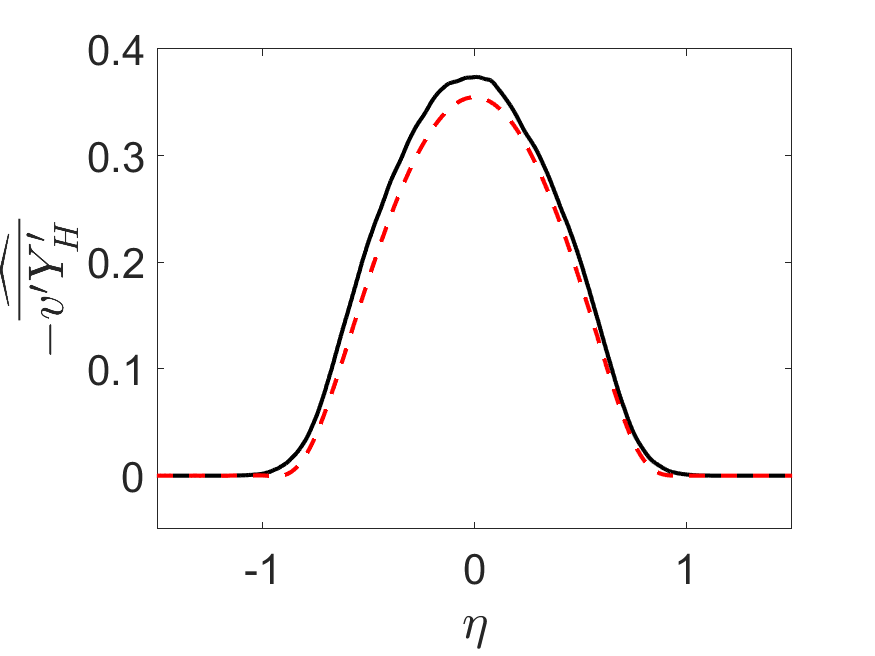}
        \subcaption[]{$D^{00}, D^{10}, D^{01}$}
    \end{subfigure}
    \caption{Turbulent scalar flux predictions using the explicit {operator} form. Captions of each plot list moments used in the model.}
    \label{fig:exp_aposteriori_tsf}
\end{figure}
\begin{figure}
    \centering
    \begin{subfigure}[]{0.49\textwidth}
        \includegraphics[width=\textwidth]{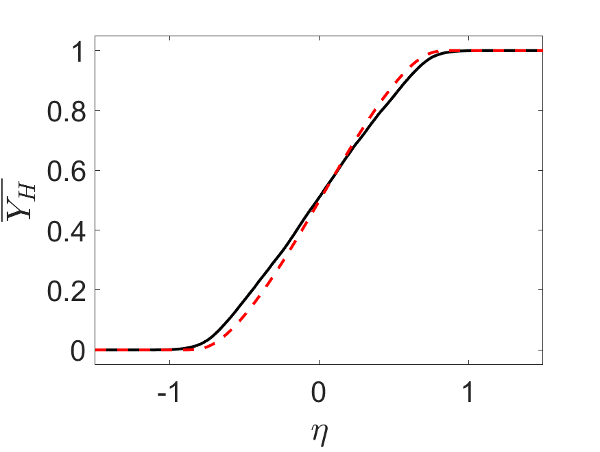}
        \subcaption[]{$D^{00}$}
    \end{subfigure}
    \begin{subfigure}[]{0.49\textwidth}
        \includegraphics[width=\textwidth]{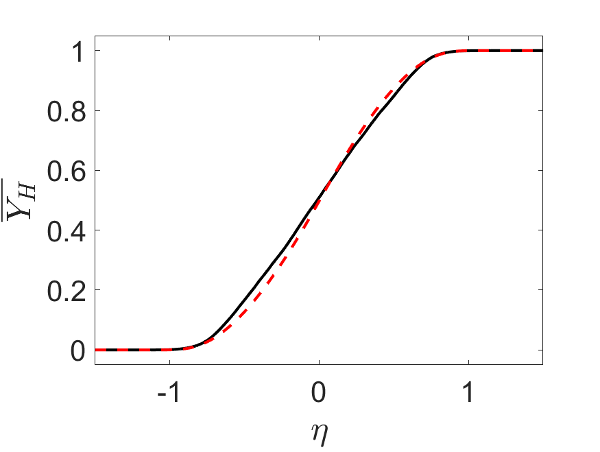}
        \subcaption[]{$D^{00}, D^{10}$}
    \end{subfigure}
    \begin{subfigure}[]{0.49\textwidth}
        \includegraphics[width=\textwidth]{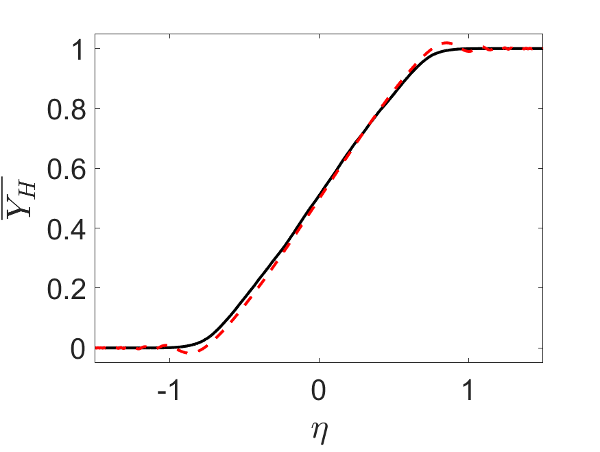}
        \subcaption[]{$D^{00}, D^{01}$}
    \end{subfigure}
    \begin{subfigure}[]{0.49\textwidth}
        \includegraphics[width=\textwidth]{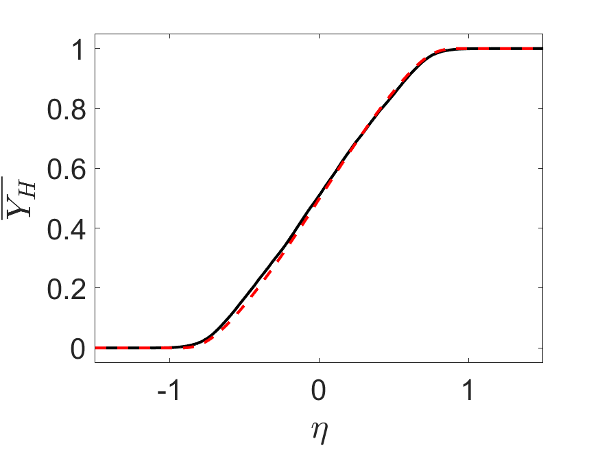}
        \subcaption[]{$D^{00}, D^{10}, D^{01}$}
    \end{subfigure}
    \caption{Mean concentration profile predictions using the explicit {operator} form. Captions of each plot list moments used in the model.}
    \label{fig:exp_aposteriori_YH}
\end{figure}

Figure \ref{fig:exp_aposteriori_tsf} shows the turbulent scalar fluxes computed using the explicit {operator} form, and figure \ref{fig:exp_aposteriori_YH} shows the corresponding mean concentration profiles.
Again, it is apparent that the leading-order moment is not enough to capture the turbulent scalar flux.
The combination using $D^{00}$, $D^{10}$, and $D^{01}$---the moments deemed most important in \S \ref{sec:term_comp}---gives the best match to the DNS measurement.

It is particularly remarkable that a converged turbulent scalar flux can be obtained using $D^{00}$, $D^{10}$, and $D^{01}$.
As mentioned previously, it is known that equation \ref{eq:explicit} may not converge.
That is, the expansion must be taken to infinite terms to remove error; truncating the expansion can result in significant error.
This is analogous to a Kramers-Moyal expansion, which cannot be approximated adequately by more than two terms, after which it requires infinite terms for convergence \citep{pawula1967,mauri1991}. 
To understand how adding terms to equation \ref{eq:explicit} can result in greater error, one can consider the eddy diffusivity kernel associated with each term.
The leading-order moment is associated with a delta function kernel, as it is purely local.
However, when equation \ref{eq:gen_eddy_diff_2d} is replaced by equation \ref{eq:explicit}, an integral operator is replaced with a high-order differential operator.
This means that the nonlocal effects are approximated by derivatives of delta functions; see \citet{liu2023} for more details.
It has been shown that, in general, the eddy diffusivity kernel is not a superposition of finite delta functions, as it is smooth \citep{manipark2021, liu2023}.
Therefore, truncation of the expansion does not match the shape of the eddy diffusivity kernel, leading to errors in prediction of the turbulent scalar flux.
{While the $D^{00}$, $D^{10}$, and $D^{01}$ combination did not diverge, adding $D^{20}$ does lead to divergent results for these reasons, so this combination is not presented here.}

Another issue with the explicit {operator} form is its numerical implementation.
In spatio-temporal space, some terms associated with higher-order moments involve mixed derivatives (e.g., the term $D^{01}\frac{\partial^2}{\partial t\partial y}$), which would undergo another spatial gradient when substituted into equation \ref{eq:ste_avg}.
Such terms are difficult to handle numerically.
In this work, the model is implemented in the more convenient self-similar space, but, ultimately, a spatio-temporal model would be developed, as it is more practical. 
It is thus pertinent to work towards a better method to incorporate nonlocal information in a RANS model that does not encounter the Kramers-Moyal-like convergence issue and is easier to implement.

\subsubsection{Implicit {operator} form and the Matched Moment Inverse (MMI)}

In this section, an implicit {operator} form is introduced as a solution to both the increasing error when adding terms from the turbulent scalar flux expansion and implementation challenges associated with the explicit {operator} form.
Recall that the explicit {operator} form fails to match the shape of the eddy diffusivity kernel without infinite terms of the turbulent scalar flux expansion.
In this implicit {operator} form, the aim is to match the shape of the eddy diffusivity kernel, instead of using the truncated expansion for the turbulent scalar flux.
Using the four moments that have been measured, this model form is
\begin{align}
    \left[1+a^{01}\frac{\partial}{\partial t}+a^{10}\frac{\partial}{\partial y}+a^{20}\frac{\partial^2}{\partial y^2}{+\dots}\right](-\langle v'Y_H' \rangle )=a^{00}\frac{\partial \langle Y_H \rangle }{\partial y},
\label{eq:imp_model_form}
\end{align}
where $a^{mn}(y,t)$ are model coefficients fitted corresponding to each of the eddy diffusivity moments $D^{mn}$ measured using MFM.
The bracketed operator on the left hand side is the Matched Moment Inverse (MMI) operator.
The way this model form is designed to match the eddy diffusivity kernel shape is detailed in \citet{liu2023}.
In addition, this form is significantly easier to implement numerically in spatio-temporal space, since it can be directly time-integrated using explicit methods.
In this way, it is also easy to add more terms with higher-order moments, as it simply requires extension of the operator.

In self-similar coordinates, this becomes
\begin{align}
    \left[1+\widehat{a^{01}}\left(1-2\eta\frac{d}{d\eta}\right)+\widehat{a^{10}}\frac{d}{d \eta}+\widehat{a^{20}}\frac{d^2}{d \eta^2}+\dots\right](-\widehat{\langle v'Y_H' \rangle } )=\widehat{a^{00}}\frac{d \langle Y_H \rangle }{d \eta},
    \label{eq:MMI_selfsim}
\end{align}
where it is found through self-similar analysis that
\begin{align}
    \widehat{a^{00}}&=\frac{1}{{\alpha^*}^2 A^2g^2(t-t^*)^3}a^{00},\\
    \widehat{a^{01}}&=\frac{1}{t-t^*}a^{01},\\
    \widehat{a^{10}}&=\frac{1}{\alpha^* Ag(t-t^*)^2}a^{10},\\
    \widehat{a^{20}}&=\frac{1}{{\alpha^*}^2 A^2g^2(t-t^*)^4}a^{20}.
\end{align}

The coefficients are determined through a process illustrated as follows in spatio-temporal coordinates for simplicity.
If one wants to construct a model in the form of equation \ref{eq:imp_model_form}, four equations must be formulated to determine the four coefficients.
This is done by using measurements from the four simulations used to determine the four moments $D^{00}$, $D^{10}$, $D^{01}$, and $D^{20}$.
For example, the first equation results from substitution of $F^{00}$ for $-\langle v'Y_H' \rangle $ and the associated desired $\frac{\partial\langle Y_H \rangle }{\partial y}$; the remaining three equations follow, using the other three moments:
\begin{align}
    \left[1+a^{10}\frac{\partial}{\partial y}+a^{01}\frac{\partial}{\partial t}+a^{20}\frac{\partial^2}{\partial y^2}\right]F^{00}&=a^{00},\\
    \left[1+a^{10}\frac{\partial}{\partial y}+a^{01}\frac{\partial}{\partial t}+a^{20}\frac{\partial^2}{\partial y^2}\right]F^{10}&=a^{00} \left(y-\frac{1}{2}\right),\\
    \left[1+a^{10}\frac{\partial}{\partial y}+a^{01}\frac{\partial}{\partial t}+a^{20}\frac{\partial^2}{\partial y^2}\right]F^{01}&=a^{00}t,\\
    \left[1+a^{10}\frac{\partial}{\partial y}+a^{01}\frac{\partial}{\partial t}+a^{20}\frac{\partial^2}{\partial y^2}\right]F^{20}&=a^{00}\frac{1}{2}\left(y-\frac{1}{2}\right)^2.
\end{align}
This system of equations is then rearranged into a matrix equation $M_\text{MMI}\mathbf{a}=\mathbf{b}$, which is solved for the coefficients in vector $\mathbf{a}=(a^{00},a^{10},a^{01},a^{20})^T$.
Note that this matrix equation is constructed over every point in space and time, so $\mathbf{a}=\mathbf{a}(y,t)$.
In this work, analysis is done in self-similar coordinates, in which $\textbf{a}=\textbf{a}(\eta)$.
If one wishes to construct a model with different moments, the MMI operator and equations must be modified accordingly.
For example, a model using only $D^{00}$ and $D^{10}$ would have an MMI operator of the form $1+a_{10}\frac{\partial}{\partial y}$ and use only the first two equations (with the $a^{01}$ and $a^{20}$ terms removed).
Thus, models using different combinations of moments would use different MMI coefficients $a^{mn}$.

To summarize, for this implicit {operator} form, the following system of equations is solved in self-similar coordinates:
\begin{align}
    \mathcal{L}_\text{MMI}\left\{-\widehat{\langle v'Y_H' \rangle } \right\}&=\widehat{a^{00}}\frac{d \langle Y_H \rangle }{d \eta},\\
    \frac{d}{d\eta}\left(-\widehat{\langle v'Y_H' \rangle }\right) &= -2\eta\frac{d\langle Y_H \rangle }{d\eta},
\end{align}
where $\mathcal{L}_\text{MMI}$ is the MMI operator constructed using some combination of moments, such as in equation \ref{eq:MMI_selfsim}.
Numerically, the following system is solved:
\begin{align}
    P(-\widehat{\langle v'Y_H' \rangle } )&=\widehat{a^{00}}\mathcal{D_\eta}\langle Y_H \rangle ,\\
    \mathcal{D_\eta}\left(-\widehat{\langle v'Y_H' \rangle }\right) &= -2\eta\mathcal{D_\eta}\langle Y_H \rangle ,
    \label{eq:apost_eqs}
\end{align}
where $P$ is the matrix representing the numerical MMI operator, and $\mathcal{D_\eta}$ is the matrix representing the numerical derivative with respect to $\eta$.
This can be rewritten as a block matrix-vector multiplication:
\begin{align}
    M \textbf{x}= 
    \begin{bmatrix}
    P & -\widehat{a^{00}}\mathcal{D_\eta}\\
    \mathcal{D_\eta} & 2\eta\mathcal{D_\eta}
    \end{bmatrix} 
    \begin{bmatrix}
    -\widehat{\langle v'Y_H' \rangle }\\
    \langle Y_H \rangle 
    \end{bmatrix}
    = \textbf{b},
\end{align}
where $\textbf{b}$ is a vector representing the right-hand side of equations \ref{eq:apost_eqs}, with the proper boundary conditions enforced.
In this study, zero gradient boundary conditions are used for the turbulent scalar flux, and Dirichlet boundary conditions are used for the mean concentration.
The system is solved using finite differences on a staggered mesh.

\begin{figure}
    \centering
    \begin{subfigure}[]{0.49\textwidth}
        \includegraphics[width=\textwidth]{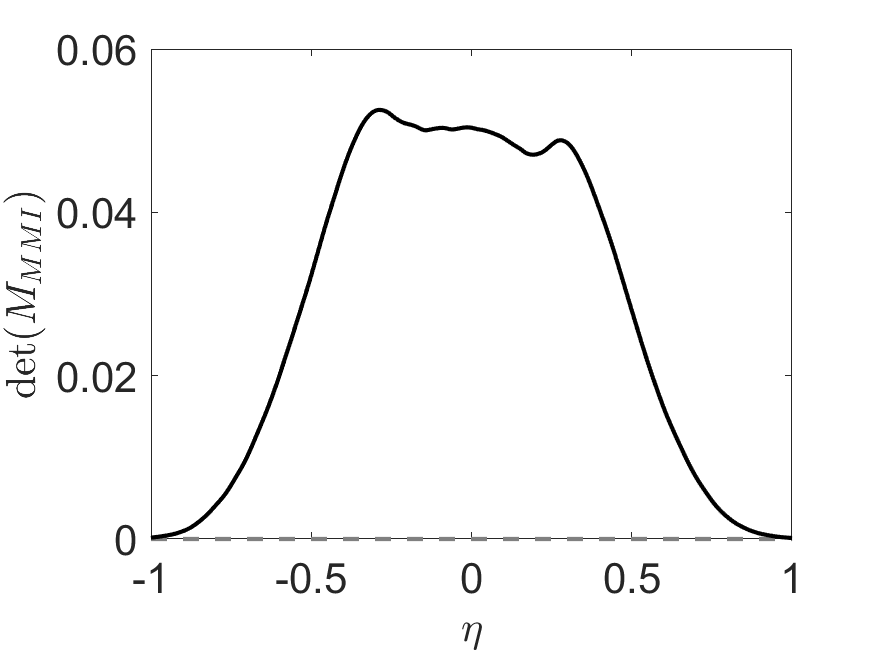}
        \subcaption[]{}
    \end{subfigure}\\
    \begin{subfigure}[]{0.49\textwidth}
        \includegraphics[width=\textwidth]{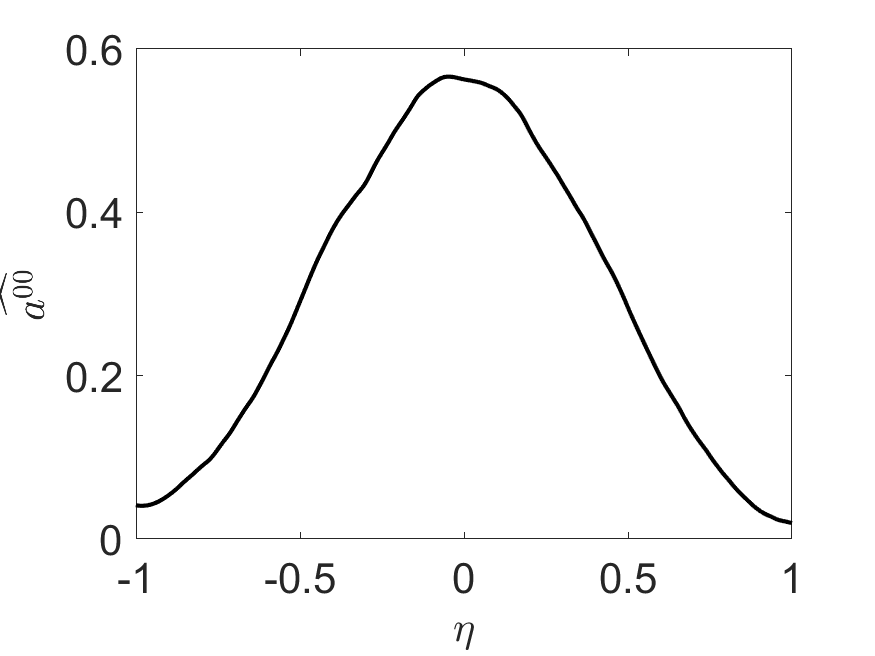}
        \subcaption[]{}
    \end{subfigure}
    \begin{subfigure}[]{0.49\textwidth}
        \includegraphics[width=\textwidth]{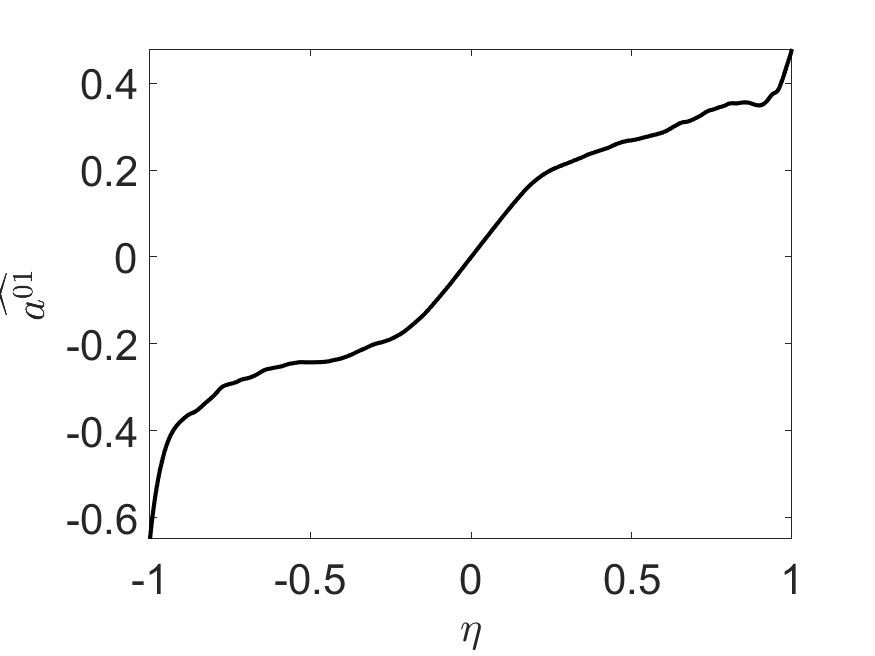}
        \subcaption[]{}
    \end{subfigure}
    \begin{subfigure}[]{0.49\textwidth}
        \includegraphics[width=\textwidth]{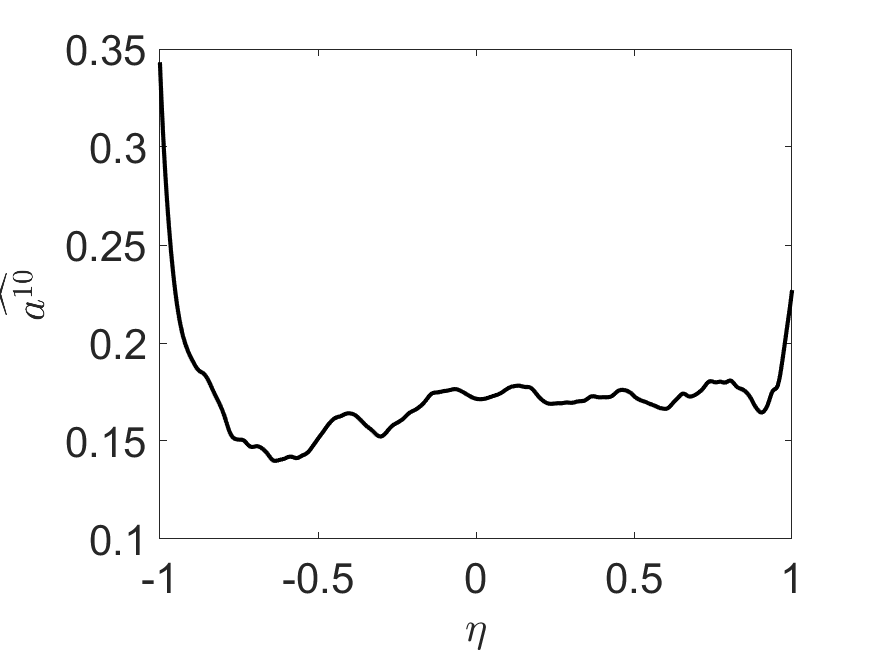}
        \subcaption[]{}
    \end{subfigure}
    \caption{(a) Determinant of $M_\text{MMI}$ over $\eta$ and (b, c, d) MMI coefficients $\widehat{a^{mn}}$ over $\eta$ for the implicit {operator} form using $D^{00}$, $D^{10}$, and $D^{01}$.}
    \label{fig:MMI_good}
\end{figure}

\begin{figure}
    \centering
    \begin{subfigure}[]{0.49\textwidth}
        \includegraphics[width=\textwidth]{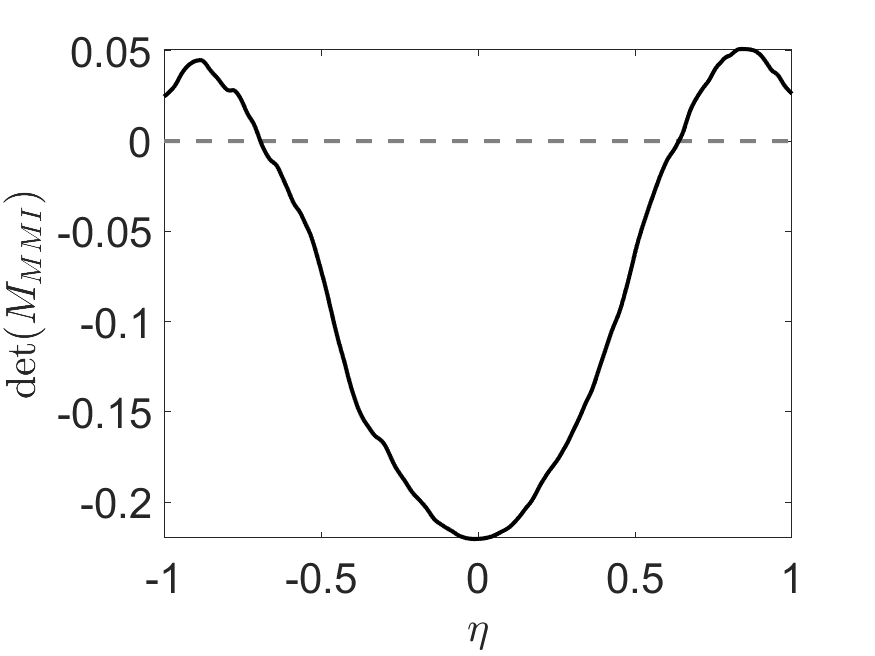}
        \subcaption[]{}
    \end{subfigure}\\
    \begin{subfigure}[]{0.49\textwidth}
        \includegraphics[width=\textwidth]{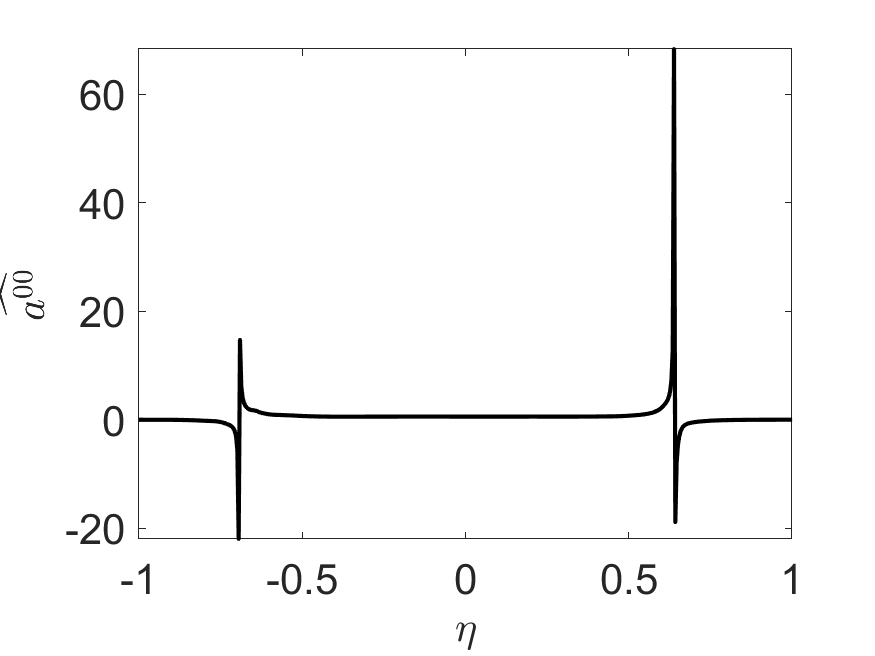}
        \subcaption[]{}
    \end{subfigure}
    \begin{subfigure}[]{0.49\textwidth}
        \includegraphics[width=\textwidth]{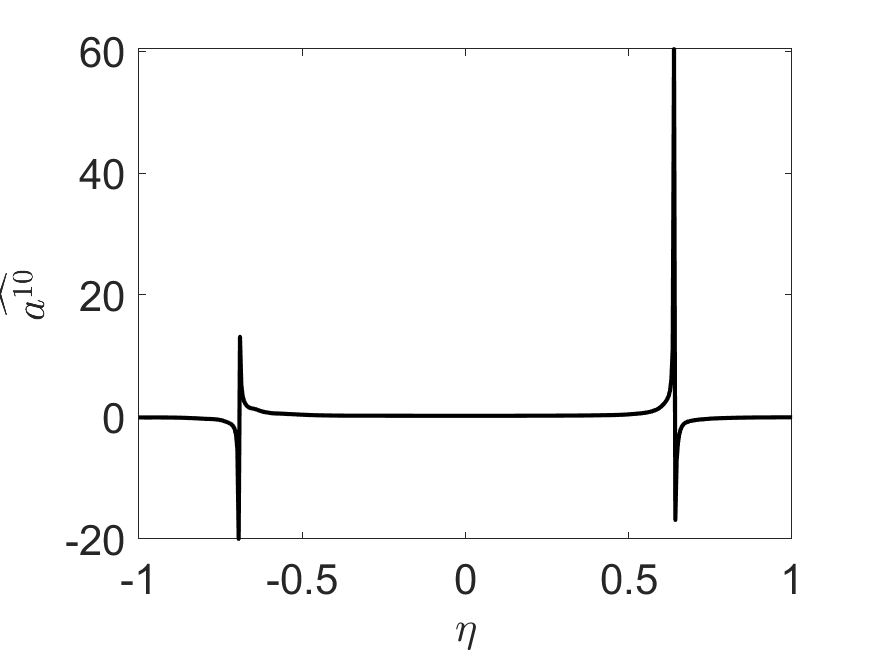}
        \subcaption[]{}
    \end{subfigure}
    \caption{(a) Determinant of $M_\text{MMI}$ over $\eta$  and (b, c) MMI coefficients $\widehat{a^{mn}}$ over $\eta$ for the implicit {operator} form using $D^{00}$ and $D^{01}$. $M_\text{MMI}$ is singular at the $\eta$ at which its determinant crosses zero.}
    \label{fig:MMI_singularity}
\end{figure}

Presented in this work are the determinants of the MMI matrix and resulting $a_{mn}$ for two different combinations of moments. Figure \ref{fig:MMI_good} shows that with the combination of $D^{00}$, $D^{10}$, and $D^{01}$, the determinant of the MMI matrix is positive for all $\eta$, so $a_{mn}$ are all well-behaved.
This is indicative of good model form.
On the other hand, figure \ref{fig:MMI_singularity} shows that with the combination of $D^{00}$ and $D^{01}$, the determinant of the MMI matrix crosses zero, so $a_{mn}$ contain singularities which effect poor RANS predictions (observable in plots presented later).
Singular matrices arising in the MMI solve for a certain form of the implicit {operator} form may indicate that form is poor, in the sense that the combination of moments does not make a good RANS model.
Since MMI appears to be sensitive to the information it takes in to determine the implicit operator coefficients, one must take special care and choose a model form that avoids this issue.
It is found that MMI determinant zero-crossings do not occur for any of the moment combinations tested in this work other than the $D^{00}$ and $D^{01}$ combination, but it may happen with combinations of other higher-order moments not measured here.

\begin{figure}
    \centering
    \begin{subfigure}[]{0.49\textwidth}
        \includegraphics[width=\textwidth]{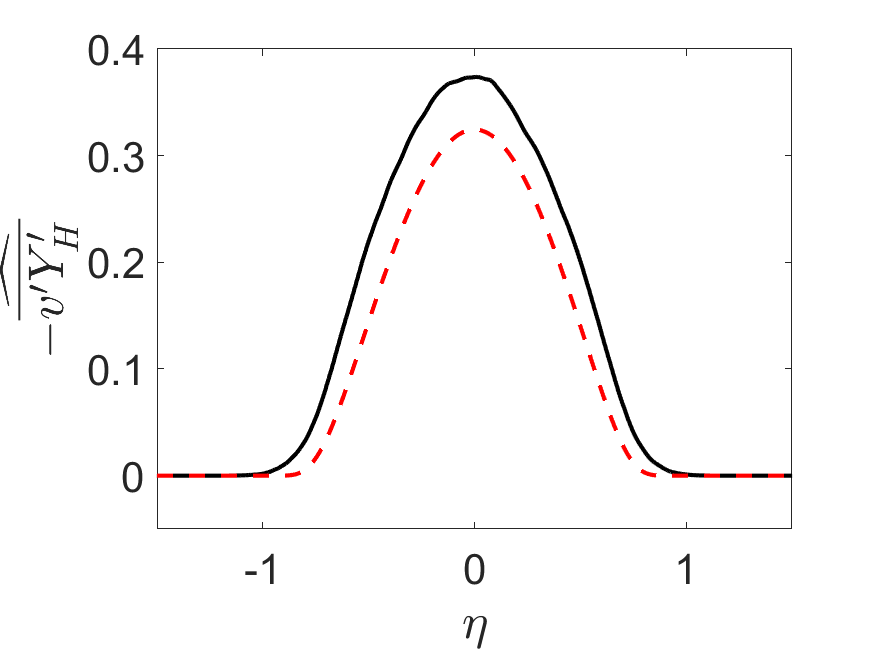}
        \subcaption[]{$D^{00}$}
    \end{subfigure}
    \begin{subfigure}[]{0.49\textwidth}
        \includegraphics[width=\textwidth]{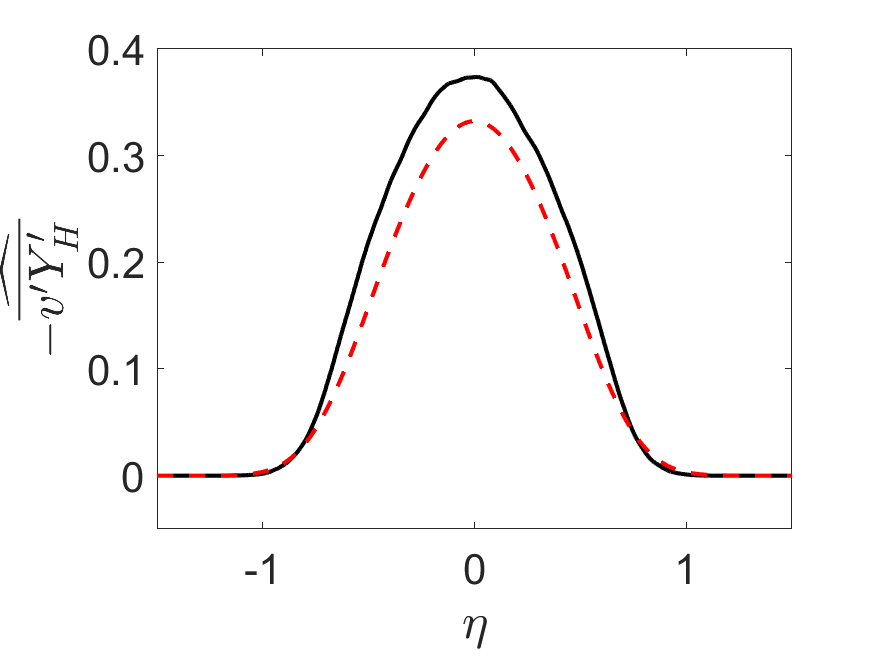}
        \subcaption[]{$D^{00}, D^{10}$}
    \end{subfigure}
    \begin{subfigure}[]{0.49\textwidth}
        \includegraphics[width=\textwidth]{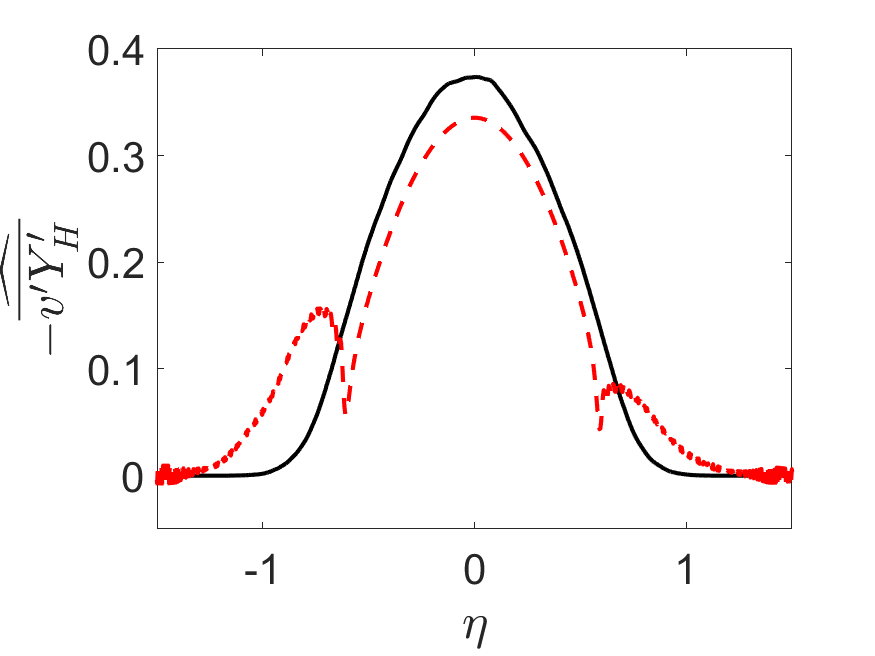}
        \subcaption[]{$D^{00}, D^{01}$}
    \end{subfigure}
    \begin{subfigure}[]{0.49\textwidth}
        \includegraphics[width=\textwidth]{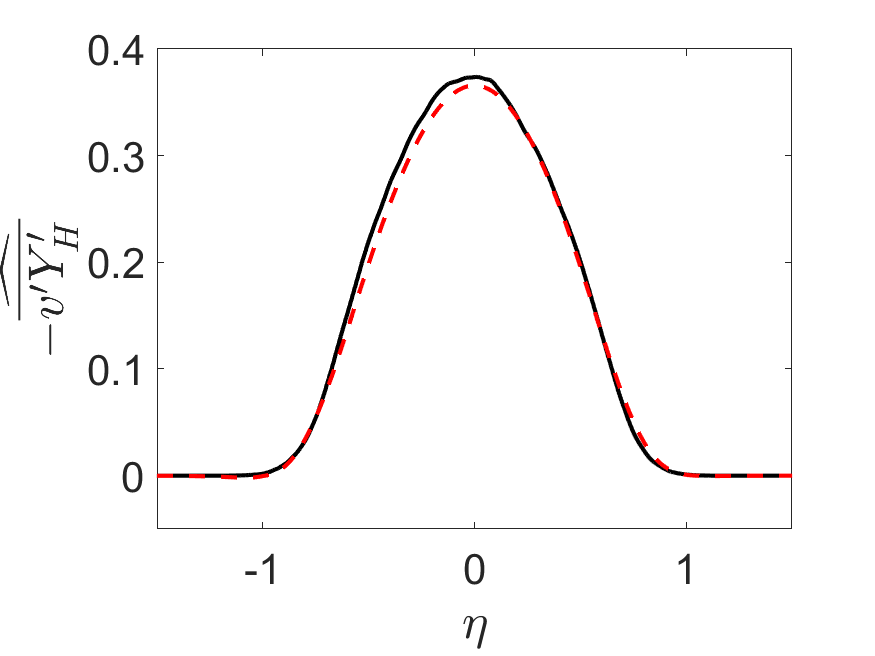}
        \subcaption[]{$D^{00}, D^{10}, D^{01}$}
    \end{subfigure}
    \caption{Turbulent scalar flux predictions using the implicit {operator} form. Captions of each plot list moments of eddy diffusivity used in the model.}
    \label{fig:imp_aposteriori_tsf}
\end{figure}

\begin{figure}
    \centering
    \begin{subfigure}[]{0.49\textwidth}
        \includegraphics[width=\textwidth]{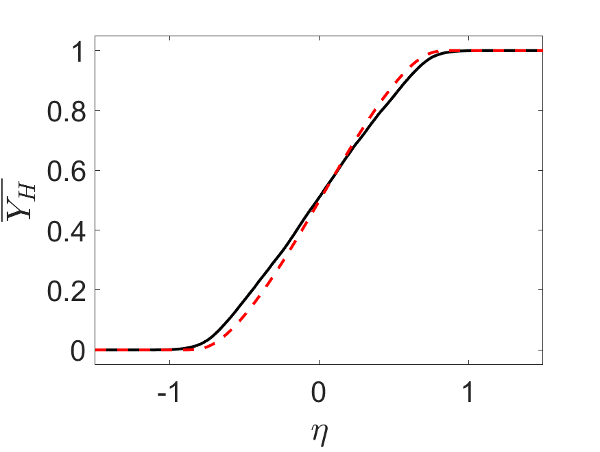}
        \subcaption[]{$D^{00}$}
    \end{subfigure}
    \begin{subfigure}[]{0.49\textwidth}
        \includegraphics[width=\textwidth]{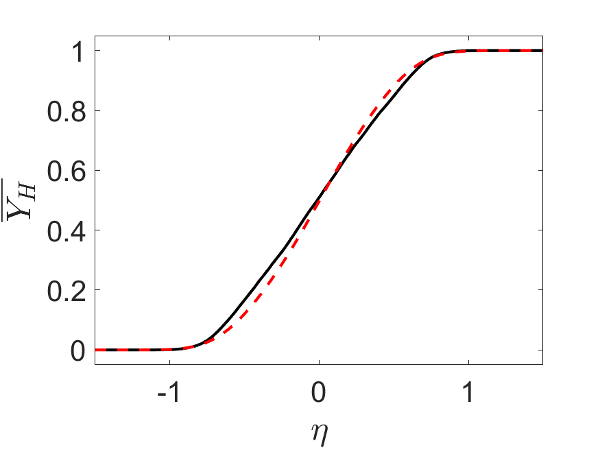}
        \subcaption[]{$D^{00}, D^{10}$}
    \end{subfigure}
    \begin{subfigure}[]{0.49\textwidth}
        \includegraphics[width=\textwidth]{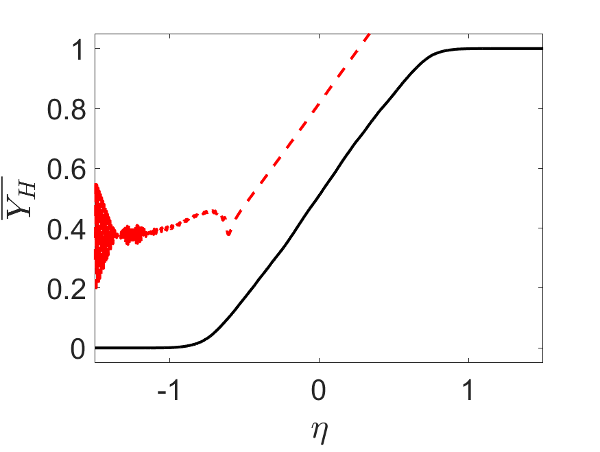}
        \subcaption[]{$D^{00}, D^{01}$}
    \end{subfigure}
    \begin{subfigure}[]{0.49\textwidth}
        \includegraphics[width=\textwidth]{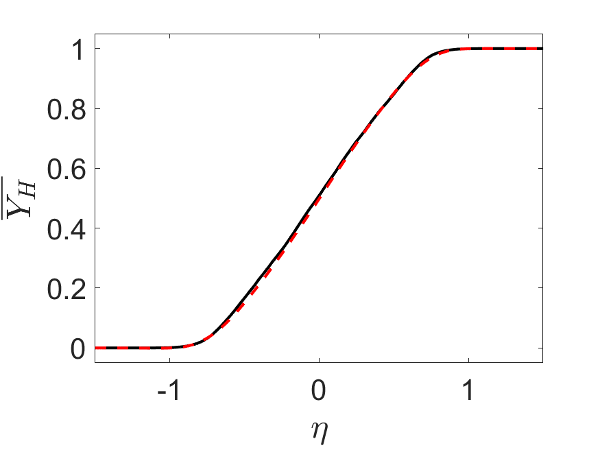}
        \subcaption[]{$D^{00}, D^{10}, D^{01}$}
    \end{subfigure}
    \caption{Mean concentration profile predictions using the implicit {operator} form. Captions of each plot list moments of eddy diffusivity used in the model.}
    \label{fig:imp_aposteriori_YH}
\end{figure}

Turbulent scalar fluxes computed using the implicit {operator} form are shown in figure \ref{fig:imp_aposteriori_tsf}.
The implicit {operator} form's turbulent scalar flux prediction using just $D^{00}$ is identical to that of the explicit {operator} form, by construction.
It is apparent that adding either $D^{10}$ or $D^{01}$ alone is insufficient.
As noted earlier, adding $D^{01}$ leads to a particularly poor prediction due to singular MMI matrices at some $\eta$.
The best match to DNS is attained using the combination of $D^{00}$, $D^{10}$, and $D^{01}$.
In fact, it is evident that the implicit {operator} form using $D^{00}$, $D^{10}$, and $D^{01}$ predicts the turbulent scalar flux more accurately than the explicit {operator} form using the same moments.
This is because the implicit {operator} form is designed to match the shape of the eddy diffusivity kernel, and 
the explicit {operator} form may not be accomplishing this.

\begin{figure}
    \centering
    \includegraphics[width=0.5\textwidth]{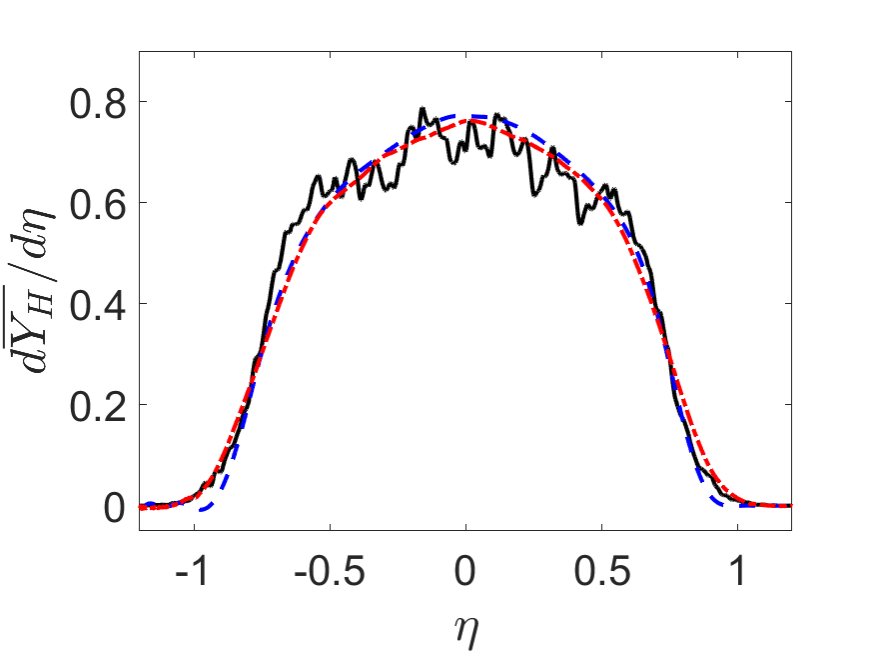}
    \caption{Derivatives of $\langle Y_H \rangle $ computed using DNS (solid black), an explicit {operator} form (dashed blue), and an implicit {operator} form (dash-dotted red).}
    \label{fig:expvsimp}
\end{figure}

These trends in the explicit and implicit {operator} forms can be observed again in the predictions of the mean concentration profile, shown in figure \ref{fig:imp_aposteriori_YH}.
Particularly, the implicit {operator} form using $D^{00}$, $D^{10}$, and $D^{01}$ gives a very good prediction of the mean concentration that nearly overlaps the DNS measurement.
For a clearer comparison of the explicit and implicit {operator} forms using these moments, figure \ref{fig:expvsimp} shows the derivatives of the DNS- and model-computed $\langle Y_H \rangle $.
The implicit {operator} form predicts a magnitude and shape closer to the DNS measurement than the explicit {operator} form does.
In particular, the implicit {operator} form captures the shape of the tails much better than the explicit {operator} form.

\section{Conclusion}

In this assessment, it is determined that nonlocality must be considered in developing more predictive models for RTI.
The studies presented in this work are facilitated using MFM, a numerical tool for precisely measuring closure operators.
Four of the eddy diffusivity moments of RTI ($D^{00}$, $D^{10}$, $D^{01}$, and $D^{20}$) are measured, and it is demonstrated that the higher-order moments, which contain information about the nonlocality of the eddy diffusivity kernel, should not be neglected when constructing models for RTI.

Specifically, it is determined that $D^{00}$, $D^{10}$, and $D^{01}$ are the most important moments for constructing a model for RTI.
Two methods for constructing RANS models using these moments are presented.
First, an explicit {operator} form, based on a Kramers-Moyal-like expansion derived by taking the Taylor series expansion of the scalar gradient in the generalized eddy diffusivity, is described and tested.
While incorporation of higher-order moments in the explicit {operator} form results in more accurate predictions than a leading-order model, there exist several issues.
One problem is that the expansion used for the explicit {operator} form may not converge, so addition of higher-order moments leads to less accurate predictions.
Another problem is that the explicit {operator} form is difficult to implement numerically.

Thus, an implicit {operator} form is presented to address these issues with the explicit {operator} form.
Since an implicit {operator} form involves an invertible matrix operator, it is easier to implement than an explicit {operator} form.
In addition, the proposed implicit {operator} form is designed to match the shape of the eddy diffusivity kernel via the MMI operator, in contrast to the explicit {operator} form, which truncates a non-converging Kramer-Moyal expansion.
It is shown that the implicit {operator} form exhibits a marked improvement in predictions over the explicit {operator} form.

{Incorporation of nonlocality into RANS models via these {operator} forms comes with several challenges.
For one, development of any new model must consider scalar realizability.
While this is not thoroughly explored in this work, since an actual model is not yet proposed, one approach to preserve realizability is suggested by \citet{braungore2021}, where the turbulent scalar flux is rewritten as an advection-like term and added to the original advection term in order to enforce physical mean component mass fractions; a conservative numerical scheme maintains realizability.
Further, the new model must be tested on more complex RTI for it to be useful in practical settings such as ICF.
This includes assessment of the model for 3D, finite Atwood, and compressible (Richtmeyer-Meshkov) flows.
The model should be tested in the same validation cases as other models for RTI, such as the tilted rig \citep{denissen2014} and gravity reversal \citep{banerjee2010}.
Based on these evaluations, which may also involve new MFM measurements where the method is extended to more complex flow regimes, the new model can be amended, as is the usual process of turbulence model development.
This is left for future work, when a new model is developed based on the findings presented here.}

One obstacle encountered in these studies is the inherent statistical error in the DNS computations.
The higher-order moments particularly contain high statistical error due to buildup of error from the lower-order moments on which they depend.
Because of this, it is admittedly difficult to draw definite conclusions about the effect of higher-order moments beyond the first-order moments.
That is, due to the statistical error, it is currently unclear if inclusion of moments beyond first-order in a RANS model would significantly improve its predictions, or if just the first-order temporal and spatial moments along with the leading-order moment are sufficient.
This motivates development of a technique to accelerate the statistical convergence of these higher-order moments.
Such a method could also be used to study the effect of other higher-order moments that were not measured in this present work, since they would have suffered from high statistical error with the current method.

{
It must be stressed that the results in this work are for 2D RTI and should not be directly applied to 3D RTI.
As noted previously, the third spatial dimension significantly impacts the turbulent physics of RTI.
In particular, 3D RTI has a lower growth rate than 2D, so lower magnitudes of the eddy diffusivity moments are expected in 3D.
Despite the quantitative difference in physics between 2D and 3D, they are qualitatively similar in the RANS space, so trends in the shapes of the eddy diffusivity moments are expected to persist in 3D.
In other words, the form of the turbulent scalar flux closure in 3D is expected to be the same as in 2D, but the coefficients would be different.
These expected trends are yet to be confirmed, and future work should involve applying MFM to 3D RTI.
}

Through this work, an understanding of nonlocality in 2D RTI has been developed.
It has been shown that incorporation of information about the nonlocality of the eddy diffusivity may greatly improve the accuracy of a RANS model.
This work demonstrates this by testing operators using MFM measurements of the nonlocal eddy diffusivity.
In practice, a RANS model for RTI would not have to rely on these MFM measurements directly; one would not have to perform many MFM simulations to construct a model.
In other words, MFM should be seen as a diagnostic tool rather than the means for building the actual model.
The ultimate goal is to develop an improved, more predictive model for RTI by incorporating nonlocal information, which the present work has demonstrated to be significant for accurate prediction of mean scalar transport in 2D RTI.

\textbf{Acknowledgements.} 
This work was performed under the auspices of the US Department of Energy by Lawrence Livermore National Laboratory under Contract No. DE-AC52-07NA27344.
D.L. was additionally supported by the Charles H. Kruger Stanford Graduate Fellowship.
J.L. was additionally supported by the Burt and Deedee McMurtry Stanford Graduate Fellowship.

\textbf{Declaration of Interests.} The authors report no conflict of interest.

\appendix
\section{Nondimensionalizations}

To determine the nondimensionalizations in equations \ref{eq:nondim1} - \ref{eq:nondim2}, a self-similarity analysis is performed.
The following self-similarity coordinate is used:
\begin{align}
    \eta = \frac{y}{{h_\text{fit}}(t)} = \frac{y}{\alpha^*Ag\left(t-t^*\right)^2}.
\end{align}
To perform transformations to this self-similar space, all derivatives are written in terms of $\eta$:
\begin{align}
    \frac{\partial}{\partial t} &= -\frac{2\eta}{t-t^*}\frac{d}{d \eta},\\
    \frac{\partial}{\partial y} &= \frac{1}{\alpha^*Ag\left(t-t^*\right)^2}\frac{d}{d \eta},\\
    \frac{\partial^2}{\partial t \partial y} &= -\frac{2}{\alpha^*Ag\left(t-t^*\right)^3}\left(\frac{\partial}{d \eta} + {\eta}\frac{d^2}{d \eta^2}\right).
\end{align}
To nondimensionalize the eddy diffusivity moments, equation \ref{eq:explicit} is substituted into equation \ref{eq:ste_avg}:
\begin{align}
    \frac{\partial \langle Y_H \rangle }{\partial t} = \frac{\partial}{\partial y}\left(D^{00}\frac{\partial \langle Y_H \rangle }{\partial y} +
    D^{10}\frac{\partial^2 \langle Y_H \rangle }{\partial y^2} +
    D^{01}\frac{\partial^2 \langle Y_H \rangle }{\partial t \partial y} +
    D^{20}\frac{\partial^3 \langle Y_H \rangle }{\partial y^3}+\hdots\right).
\end{align}
The equation is then transformed to self-similar space:
\begin{align}
    -\frac{2\eta}{t-t^*}\frac{d \langle Y_H \rangle }{d \eta} =
    \frac{1}{\alpha^*Ag\left(t-t^*\right)^2}\frac{d}{d \eta}&\left[
    \frac{1}{\alpha^*Ag\left(t-t^*\right)^2} D^{00}\frac{d \langle Y_H \rangle }{d \eta} \right.\nonumber\\
    &+\frac{1}{{\alpha^*}^2A^2g^2\left(t-t^*\right)^4} D^{10}\frac{d^2 \langle Y_H \rangle }{d \eta^2} \nonumber\\
    &-\frac{2}{\alpha^*Ag\left(t-t^*\right)^3} D^{01}\left(\frac{d \langle Y_H \rangle }{d \eta} + \eta\frac{d^2 \langle Y_H \rangle }{d \eta^2}\right) \nonumber\\
    &\left.+\frac{1}{{\alpha^*}^3A^3g^3\left(t-t^*\right)^6} D^{20}\frac{d^3 \langle Y_H \rangle }{d \eta^3}\hdots
    \right].
\end{align}
Rearranging,
\begin{align}
    -2\eta\frac{d \langle Y_H \rangle }{d \eta} = \frac{d}{d \eta}&\left[
    \frac{1}{{\alpha^*}^2A^2g^2\left(t-t^*\right)^3} D^{00}\frac{d \langle Y_H \rangle }{d \eta} \right.\nonumber\\
    &+\frac{1}{{\alpha^*}^3A^3g^3\left(t-t^*\right)^5} D^{10}\frac{d^2 \langle Y_H \rangle }{d \eta^2} \nonumber\\
    &-\frac{2}{{\alpha^*}^2A^2g^2\left(t-t^*\right)^4} D^{01}\left(\frac{d \langle Y_H \rangle }{d \eta} + \eta\frac{d^2 \langle Y_H \rangle }{d \eta^2}\right) \nonumber\\
    &\left.+\frac{1}{{\alpha^*}^4A^4g^4\left(t-t^*\right)^7} D^{20}\frac{d^3 \langle Y_H \rangle }{d \eta^3}\hdots
    \right].
\end{align}
This reveals nondimensionalizations for the eddy diffusivity moments.
The prefactors to the derivatives of $\langle Y_H \rangle $ on the right hand side are denoted as the normalized eddy diffusivity moments $\widehat{D^{mn}}$.

The turbulent scalar flux scales with the leading-order term in equation \ref{eq:explicit}.
Substitution of the nondimensionalization for $D^{00}$ (equation \ref{eq:nondimD00}) into the leading-order term in equation \ref{eq:explicit} and transformation to self-similar coordinates gives the scaling for the turbulent scalar flux:
\begin{align}
    -\langle v'Y_H' \rangle  \sim {\alpha^*}Ag\left(t-t^*\right)\widehat{D^{00}}\frac{d\langle Y_H \rangle }{d \eta}.
\end{align}

\begin{figure}
    \centering
    \begin{subfigure}[]{0.49\textwidth}
        \includegraphics[width=\textwidth]{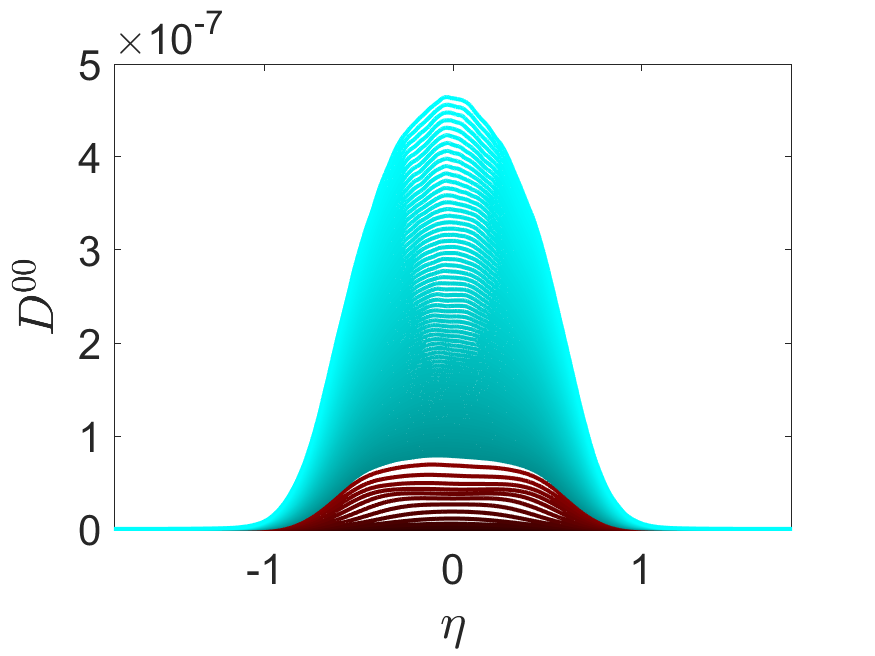}
        \subcaption[]{}
    \end{subfigure}
    \begin{subfigure}[]{0.49\textwidth}
        \includegraphics[width=\textwidth]{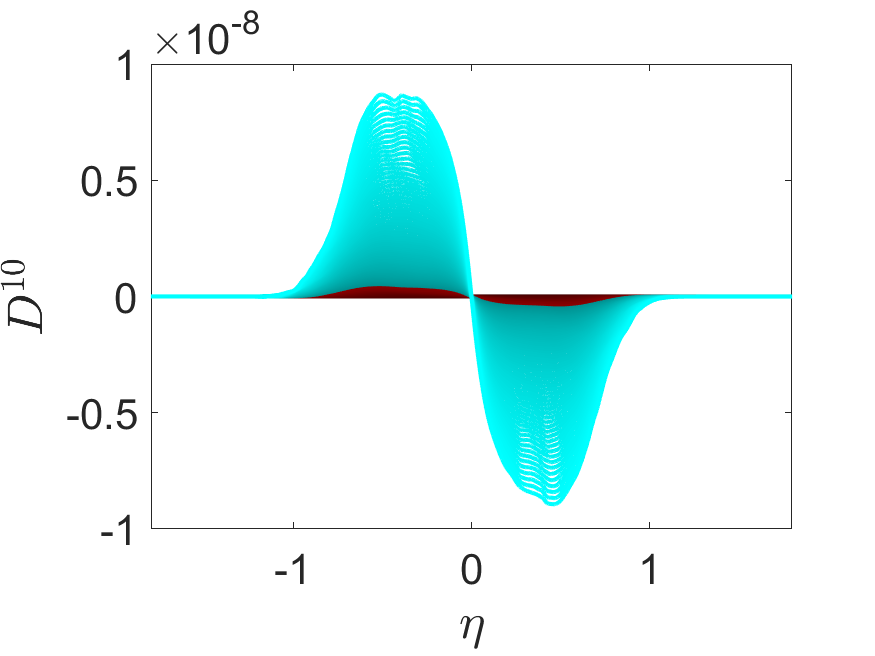}
        \subcaption[]{}
    \end{subfigure}
    \begin{subfigure}[]{0.49\textwidth}
        \includegraphics[width=\textwidth]{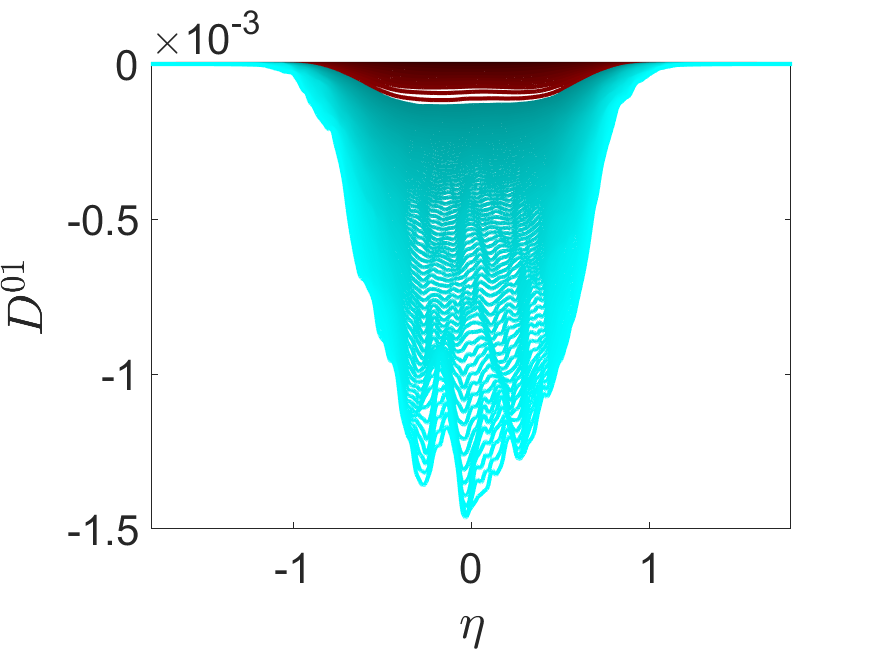}
        \subcaption[]{}
    \end{subfigure}
    \begin{subfigure}[]{0.49\textwidth}
        \includegraphics[width=\textwidth]{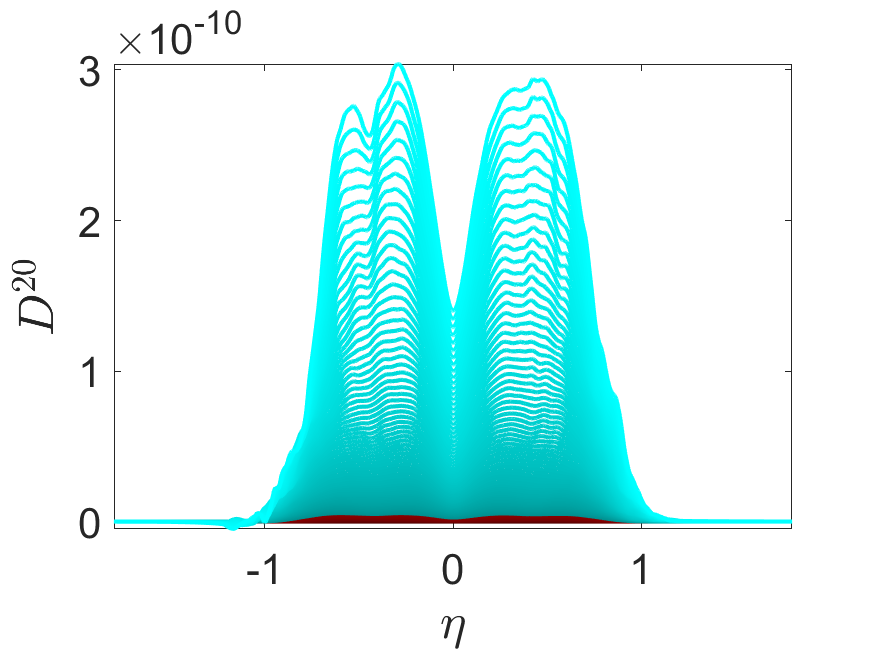}
        \subcaption[]{}
    \end{subfigure}
    \caption{Unscaled moments at different times over $x$. Dimensions of the moments are as follows: $D^{00}$ is $m^2/s$, $D^{10}$ is $m^3/s$, $D^{01}$ is $m^2$, and $D^{20}$ is $m^4/s$. Red profiles are from early times, and cyan profiles are from late times, when the flow is self-similar. Lighter lines correspond to later times; darker lines correspond to earlier times.}
    \label{fig:raw_moms}
\end{figure}

\begin{figure}
    \centering
    \begin{subfigure}[]{0.49\textwidth}
        \includegraphics[width=\textwidth]{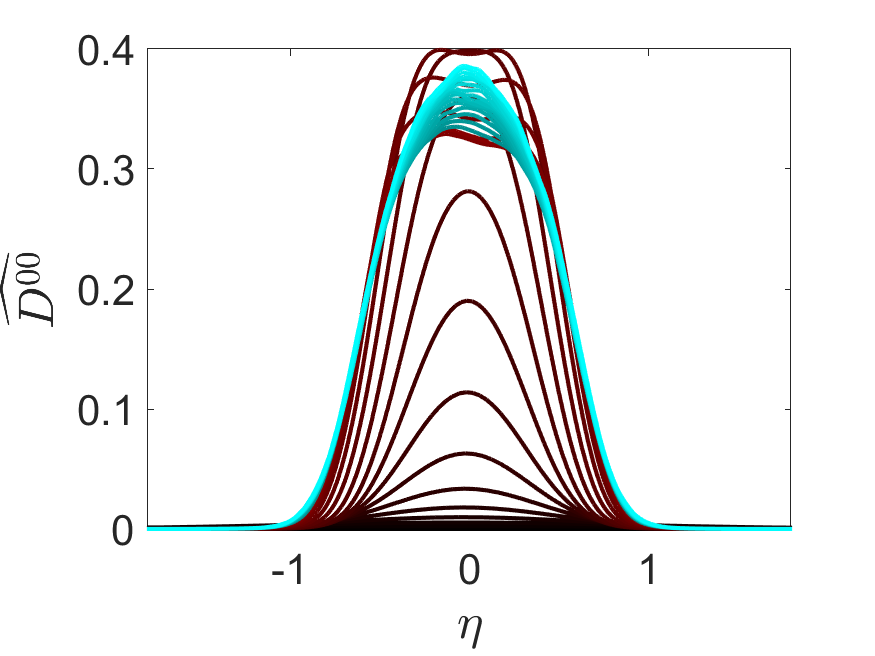}
        \subcaption[]{}
    \end{subfigure}
    \begin{subfigure}[]{0.49\textwidth}
        \includegraphics[width=\textwidth]{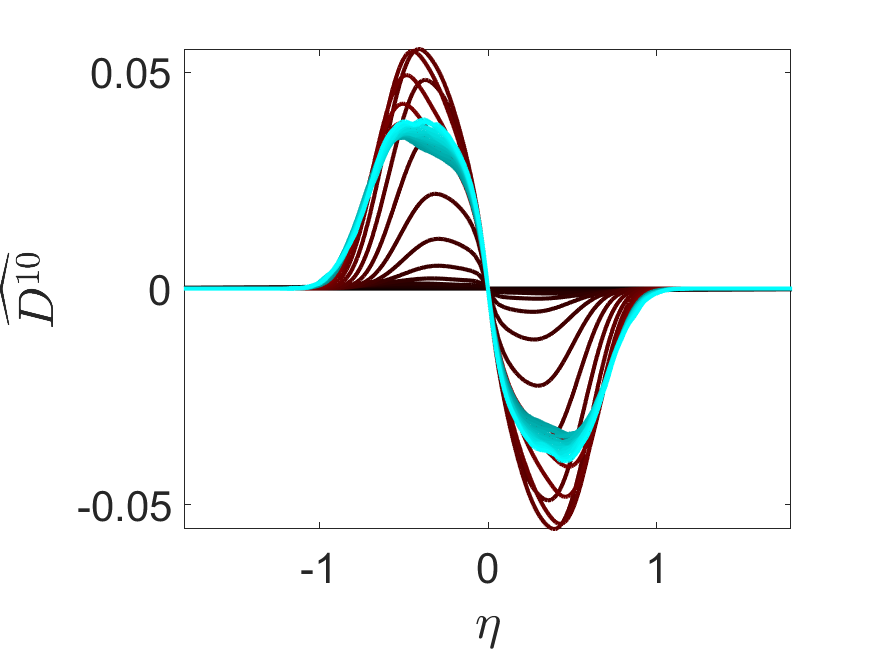}
        \subcaption[]{}
    \end{subfigure}
    \begin{subfigure}[]{0.49\textwidth}
        \includegraphics[width=\textwidth]{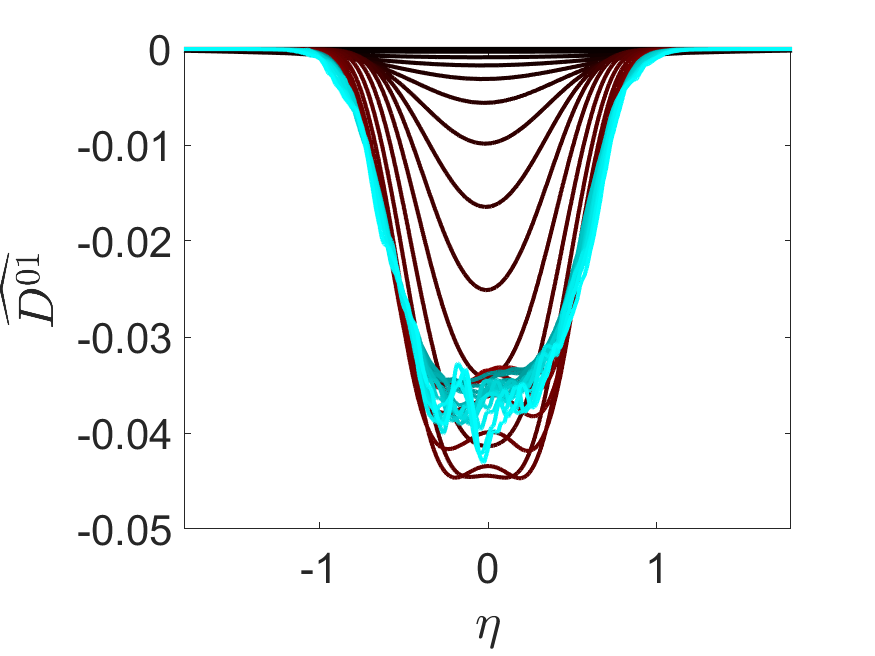}
        \subcaption[]{}
    \end{subfigure}
    \begin{subfigure}[]{0.49\textwidth}
        \includegraphics[width=\textwidth]{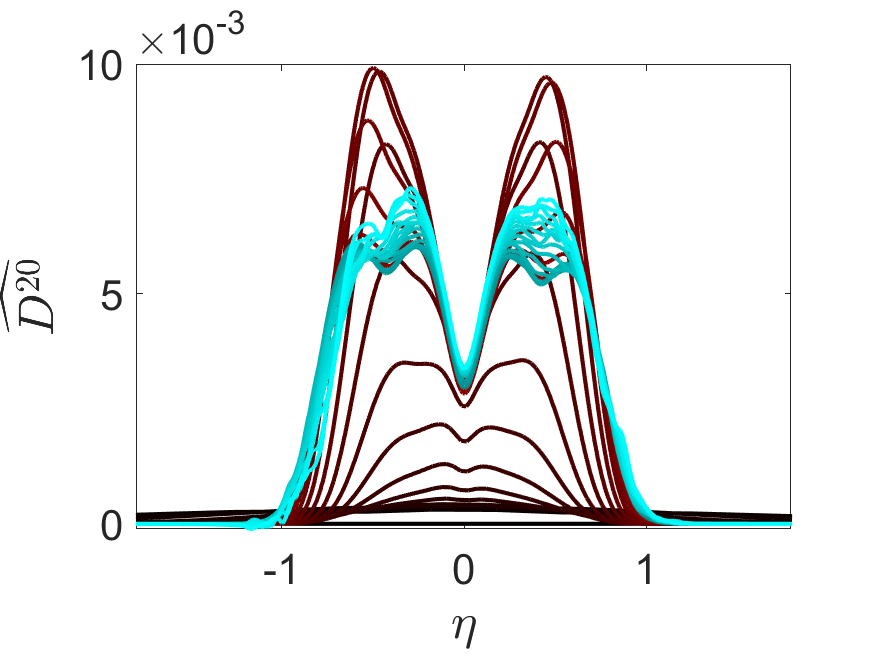}
        \subcaption[]{}
    \end{subfigure}
    \caption{Self-similar collapse of moments. All $\widehat{D^{mn}}$ are dimensionless. Red profiles are from early times, and cyan profiles are from late times, when the flow is self-similar. Lighter lines correspond to later times; darker lines correspond to earlier times.}
    \label{fig:selfsim_moms}
\end{figure}

\begin{figure}
    \centering
    \includegraphics[width=0.5\textwidth]{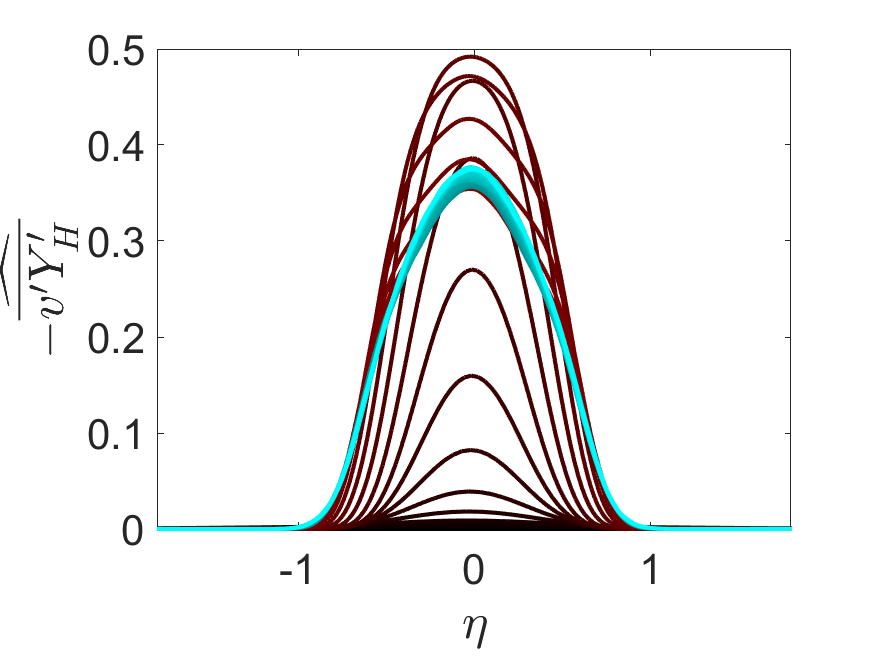}
    \caption{Self-similar collapse of turbulent scalar flux. $-\widehat{\langle v'Y_H' \rangle }$ is dimensionless. Red profiles are from early times, and cyan profiles are from late times, when the flow is self-similar. Lighter lines correspond to later times; darker lines correspond to earlier times.}
    \label{fig:selfsim_tsf}
\end{figure}

\begin{figure}
    \centering
    \includegraphics[width=0.5\textwidth]{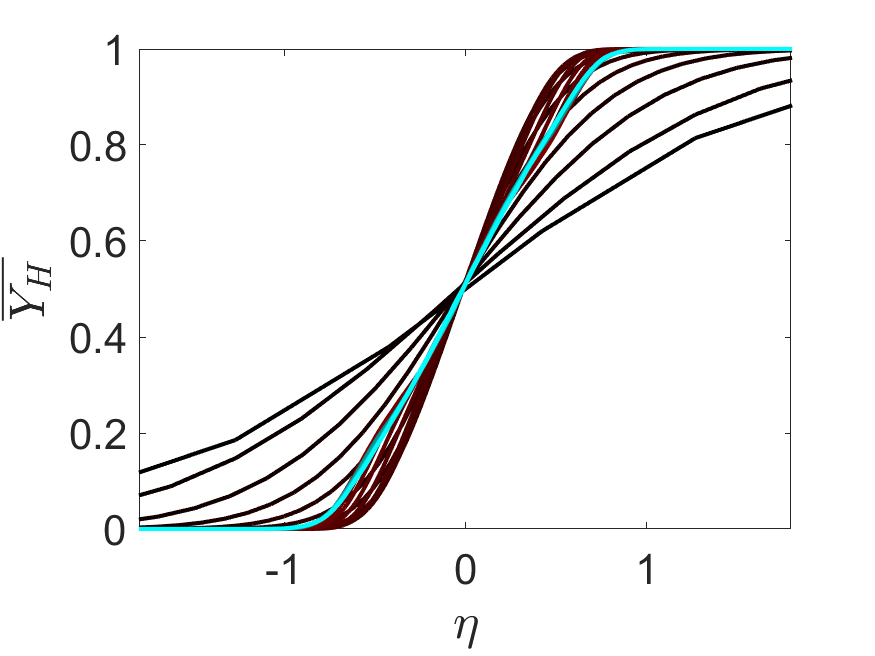}
    \caption{ Mean concentration profiles at different times. Red profiles are from early times, and cyan profiles are from late times, when the flow is self-similar. Lighter lines correspond to later times; darker lines correspond to earlier times.}
    \label{fig:selfsim_yh}
\end{figure}
Figure \ref{fig:raw_moms} shows the unscaled moments measured directly from the MFM simulations.
The profiles are taken from the portion of the simulation where the flow is self-similar ($\tau\gtrapprox17$).
It is obvious that without normalizing the moments as described above there is no self-similar collapse.
The moments are scaled and plotted against $\eta$ in figure \ref{fig:selfsim_moms} to demonstrate the self-similar collapse.
The normalized turbulent scalar flux and mean concentration profiles are shown in figures \ref{fig:selfsim_tsf} and \ref{fig:selfsim_yh}, also showing self-similar collapse.

\bibliographystyle{jfm}
\bibliography{jfm-instructions}
    
\end{document}